\documentclass[linenumbers, astrosymb, tighten, fleqn]{cas-dc}

\usepackage{amsmath}
\usepackage{lineno}
\usepackage{mathtools}
\usepackage{url}
\usepackage{gensymb}
\usepackage{subcaption}
\usepackage{setspace}

\usepackage{cancel}
\usepackage{xspace}
\usepackage{xfrac}
\newcommand{\de}{{\rm d}}
\newcommand{\dasp}{\alpha_d}
\newcommand{\fL}{f_L}

\newcommand{\DIAD}{\texttt{DIAD}\xspace}
\newcommand{\RAD}{\texttt{RADMC-3D}\xspace}
\newcommand{\RADp}{\texttt{RAD+}\xspace}

\graphicspath{{./}{figures/}}

\usepackage[range-phrase=-, range-units=single, list-final-separator=\text{, and }]{siunitx}
\usepackage[authoryear]{natbib}
\begin{document}

\title[mode=title]{A Retrieval Framework for Observationally Constraining the Parameters of Circumplanetary Disks}

\shorttitle{Constraining CPD Parameters}
\shortauthors{Taylor \& Adams}

\author[1]{Aster G. Taylor}[orcid=0000-0002-0140-4475]
\cormark[1]
\fnmark[1]
\ead{agtaylor@umich.edu}
\affiliation[1]{organization={Department of Astronomy, University of Michigan},
    % addressline={}, 
    city={Ann Arbor},
    % citysep={}, % Uncomment if no comma needed between city and postcode
    postcode={48109}, 
    state={MI},
    country={USA}}

\author[2,1]{Fred C. Adams}[orcid=0000-0002-8167-1767]
\affiliation[2]{organization={Department of Physics, University of Michigan},
    % addressline={}, 
    city={Ann Arbor},
    % citysep={}, % Uncomment if no comma needed between city and postcode
    postcode={48109}, 
    state={MI},
    country={USA}}

\cortext[cor1]{Corresponding author}
\fntext[fn1]{Fannie and John Hertz Foundation Fellow}

% \author[0000-0002-3950-5386]{Nuria Calvet}
% \affiliation{Dept. of Astronomy, University of Michigan, Ann Arbor, MI 48109, USA}
% \email{ncalvet@umich.edu}

\begin{abstract}
\normalsize
As they form, giant planets are surrounded by disks of gas and dust sourced from the background circumstellar disk. 
Although there have been few detections to date, upcoming instruments are likely to discover many more of these systems in the coming decades. 
Accurate spectral modeling will enable these observations to constrain the properties of these forming systems. 
Towards this end, we have constructed a semianalytic model for the structure and radiative signatures of geometrically thick circumplanetary disks and their planet hosts. 
Fitting these radiative signatures to synthetic observations of a two-dimensional disk model then quantifies the parameter constraints that can be derived {(subject to model assumptions)}. 
This machinery provides estimates of the values and uncertainties in system parameters, and some combinations of parameters have significantly smaller uncertainties than others. 
This model is then used to fit observations of real protoplanets, with good results. 
The derived parameters provide useful context about the local extinction, formation history, and initial entropy of these objects. 
\end{abstract}


\maketitle

\section{Introduction}

% Since the detection of the first exoplanet more than thirty years ago \citep{Mayor1995}, more than \num{6000} exoplanets have now been discovered. 
%While understanding of the structure and composition of giant planets has made significant progress, the origins of these exoplanets remain under study. Of particular interest is the formation processes of giant planets, which exert a significant influence on their planetary systems. 

Although giant planets are now routinely detected in orbit around other stars, the origins of these exoplanets remains under study. 
Most giant planets are expected to form through the core accretion paradigm \citep{Pollack1996}.
This process can be divided into three stages. 
In the first stage, solid material in the circumstellar disk coagulates through some mechanism (streaming instability, pebble accretion, etc.) into a $\sim\qty{10}{M_\oplus}$ core. 

This core then begins to dominate its local environment and accretes gaseous material from the background disk into a hydrostatically-supported envelope that fills the Hill radius. 
As this envelope cools and contracts, more material is added until the protoplanet reaches a mass of $\sim\qty{20}{M_\oplus}$. 
Beyond this mass, the envelope can no longer be fully supported by the hydrostatic pressure and contracts rapidly. 
The protoplanet then begins a phase of runaway mass accretion through the Hill sphere. 
This third and final phase sets most of the mass of the planet and is thus critical for understanding the formation process. 

In this phase, the infalling material is expected to form a circumplanetary disk (see, e.g., \citealt{Lubow1999, Canup2002}). 
Much like the formation of a protostellar disk, conservation of angular momentum prevents the accretion flow from collapsing directly to the planet. 
Instead, most of the infalling material shocks near the midplane, dissipating its vertical velocity and forming a circumplanetary disk. 
The disk transfers angular momentum and processes most of its material onto the planetary surface. 
Most of the mass that eventually makes up the giant planet is accreted during this stage.

Although this overall picture is relatively well-accepted, significant uncertainty remains. 
While hydrodynamic simulations show that these disks form (e.g., \citealt{Szulagyi2016, Lambrechts2019, Lega2024} and many others), there are significant qualitative differences depending on model assumptions, the gas equation of state, and the included physics. 
For example, some work (e.g., \citealt{Ayliffe2009}) shows that the material is accreted primarily through the equator of the system, others (e.g., {\citealt{Lambrechts2017, Sagynbayeva2025}} ) show infall primarily along the system poles, and some work shows accretion primarily through the mid-latitudes \citep{Li2023}. 
There are also significant disagreements on how the accretion rate depends on the mass (e.g., \citealt{Maeda2022, Choksi2023, Krapp2024}).
These distinctions have significant implications for the masses and compositions of giant planets. 

Although observing these systems during the formation process provides the best opportunity to resolve these differences, only a few forming giant planets and their circumplanetary disks have been observed to date \citep{Isella2019, Benisty2021, Christiaens2024a, vanCapelleveen2025, Fasano2025}. 
In the coming years, \textit{JWST} and the Extremely Large Telescope (ELT) are expected to detect and characterize more of these forming systems. 
Since the circumplanetary environment has a typical size of $\sim\qty{0.3}{au}$ for a planet with a mass of \qty{1}{M_J} at \qty{5}{au}, even the next generation of instruments will not have sufficient angular resolution to spatially resolve the protoplanetary structure. 

Models of the spectral energy distributions (SEDs) of circumplanetary environment are thus necessary to interpret the expected observations. Although previous work has constructed semianalytic models of the structure and radiative signatures of circumplanetary disks {(e.g., \citealt{Adams2022, Taylor2024, Sun2024, Taylor2025, Choksi2025, Sun2026})}, these models assume  {(either implicitly or explicitly)} that the disk is geometrically thin and optically thick.
However, numerical models of these systems show that circumplanetary disks are {in fact} geometrically thick \citep{Szulagyi2016, Choksi2025, Taylor2026}.
This geometry affects the structure and the emergent radiative signatures of the forming planets and their disks and must be included in state-of-the-art models.

This work constructs a semianalytic model for the structure and radiative signatures of geometrically thick circumplanetary disks and investigates how this model can constrain the parameters of observed protoplanets. 
The model is calibrated to accurately match the structure calculated by previous numerical work, which surveyed a wide range of the parameter space. 

This paper is structured as follows. The semianalytic model and the numerical model are briefly described and validated in Sec. \ref{sec:SAM}.
In Sec. \ref{sec:synthobv}, a Markov Chain Monte Carlo (MCMC; \citealt{Metropolis1953}) is used to fit this model to synthetic SEDs in a variety of wavelength bands. 
These results provide concrete statistical measures of how much information can be derived from observations in a given wavelength band and in a particular system. 
The model is then fit to the SEDs of the PDS 70 protoplanets and GQ Lup b in Sec. \ref{sec:realplanets}. 
The paper concludes in Sec. \ref{sec:conc} with a summary of our results and a discussion of their implications.

\begin{figure*}[t]
    \centering
    \includegraphics[width=\linewidth]{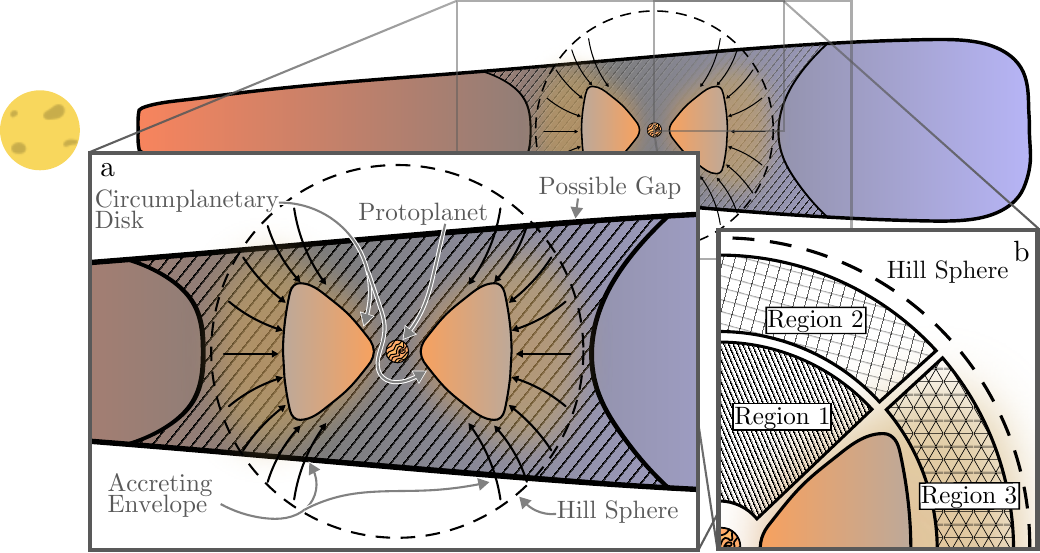}
    \caption{\textbf{System Schematic.} A schematic of the circumplanetary system, shown as a meridional cross-section. This figure is not to scale. The planet orbits the star within a circumstellar disk, which may or may not have a gap at the planet's location (hashed region). The Hill sphere forms the boundary between the circumstellar disk and the circumplanetary environment. Material is accreted through the Hill sphere onto the circumplanetary disk, which then processes this material onto the central planet. Subpanel a shows a zoomed-in view of the circumplanetary environment. Subpanel b shows the circumplanetary environment along with the regions of the envelope. Region 1, between the disk and the pole, is heated directly by the planet and can only be seen by an observer over a narrow range of viewing angles. Region 2 is also heated by the planet but can be observed from any viewing geometry. Region 3 is heated only by the outer edge of the disk and also can be seen from any viewing angle.}
    \vspace{-10pt}
    \label{fig:cpdgeom}
\end{figure*}

\setlength{\tabcolsep}{2pt}
\begin{table}
    \caption{\textbf{Canonical Fiducial Values.} Fiducial values for model parameters, which are repeatedly used throughout this paper. }
    \centering
    \begin{tabular}[t]{rcl}
        Variable & Symbol & Default Value \\\hline
        Stellar mass & $M_\star$ & \qty{1}{M_\odot} \\
        Orbital distance & $a$ & \qty{5}{au} \\
        Mass accretion rate & $\dot{M}$ & \qty{10}{M_J\per\mega yr} \\
        Viewing angle & $\psi$ & 0; polar\\
        Planet mass & $M_p$ & \qty{1}{M_J} \\
        Planet radius & $R_p$ & \qty{1.4}{R_J} \\
        Planet magnetic field & $B_{p,0}$ & \qty{500}{G} \\
        Infall geometry & - & isotropic \\
        Silicate fraction & $f_{\rm Si}$ & \num{1.0} \\
        Dust-to-gas ratio & $\eta$ & \num{0.0065} \\
        Viscosity parameter & $\alpha$ & 0.1 \\
        Angular momentum deficit & $\lambda_d$ & 1.0 \\
    \end{tabular}
    \label{tab:canonvals}
\end{table}

\section{Model Construction}\label{sec:SAM}

This section describes of the circumplanetary system and the methods used to calculate its structure and radiative signatures. These models are focused on a central protoplanet, which is surrounded by a circumplanetary disk. The infalling material, which feeds the disk and the planet, constitutes the circumplanetary envelope. The entire planet/disk/envelope system orbits a central host star within the star's natal circumstellar disk. If the planet is sufficiently massive or the circumstellar disk is thin, then the planet can clear a gap in the circumstellar disk. This setup is shown schematically in Fig. \ref{fig:cpdgeom}. The circumplanetary system is described by a set of system parameters (e.g., planet mass, mass accretion rate, orbital distance, etc.). These parameters, along with their default values, are listed in Table \ref{tab:canonvals}. 

The boundary between the circumstellar disk and the protoplanetary environment is here set to be the Hill surface. This surface is assumed to be a sphere with a radius $R_H=a(M_p/3M_\star)^{1/3}$. The circumplanetary system is fed through the Hill sphere by the background circumstellar disk at a mass accretion rate $\dot{M}$. Because some material that enters this surface is expected to subsequently leave the Hill sphere (e.g., \citealt{Ayliffe2009, Lambrechts2017, Szulagyi2016}), $\dot{M}$ is defined here to be the net mass accretion rate onto the planet. In this work, the angular distribution of the infalling material is modeled as isotropic.\footnote{Previous work has included polar asymmetries in the distribution of infalling material (e.g.,\citealt{Taylor2024}). This model still has this capacity, but observations cannot distinguish between different infall functions. }
% Following previous work \citep{Taylor2024}, the infalling material is characterized by a polar asymmetry function $f_i$. This function is defined such that the net mass accretion rate through the Hill sphere from a polar angle $\theta_0$ is given by $\dot{M}f_i(\theta_0)$. For example, these functions take the form 
% \begin{equation}
%     f_i(\theta_0)=\begin{cases}
%     3\cos^2\theta_0 & \text{polar;}\\
%     1 & \text{isotropic;}\\
%     \tfrac{3}{2}(1-\cos^2\theta_0) & \text{equatorial.}
%     \end{cases}
% \end{equation}

The infall is assumed to be ballistic so that mass is conserved along streamlines, fully specifying the density of the envelope. The form of the density has been given many times (e.g., \citealt{Ulrich1976, Cassen1981, Chevalier1983}, etc.) and will not be repeated here. 

Due to conservation of angular momentum, the infalling material shocks at the midplane and forms a circumplanetary disk (see, e.g., \citealt{Machida2008, Szulagyi2017, Fung2019}). This disk has an outer boundary given by the centrifugal radius \citep{Quillen1998, Martin2011, Tanigawa2012}, which takes the form
\begin{equation}
    R_C=\frac{a}{3}\left(\frac{M_p}{3M_\star}\right)^{1/3}\,.
\end{equation}
If the accretion flow has an angular momentum deficit characterized by $\lambda_d$, then the outer radius of the disk is reduced by a factor of $\lambda_d^2$ \citep{Adams2025}. 

This disk then viscously processes the material inwards towards the central planet. The planetary magnetic field truncates the disk at a radius $R_X$ and feed the planet along the magnetic field lines. This truncation radius is approximately given by \citep{Ghosh1978, Blandford1982}
\begin{equation}
\begin{split}
    \frac{R_X}{R_p}\simeq\,&3.8\left(\frac{M_p}{\unit{M_J}}\right)^{-1/7}\left(\frac{\dot{M}}{\qty{10}{M_J/Myr}}\right)^{-2/7}\\
    \times&\left(\frac{B_p}{\qty{500}{G}}\right)^{4/7}\left(\frac{R_p}{\qty{e10}{cm}}\right)^{5/7}\,.
\end{split}
\end{equation}
Here, $R_p$ is the radius of the planet and 
\begin{equation}
B_p=B_{p,0}\left(\frac{\dot{M}}{\qty{10}{M_J/Myr}}\right)^{1/3}
\end{equation}
is the planetary surface magnetic field strength. Although in practice there is significant scatter between $R_X$ and $\dot{M}$ \citep{Thanathibodee2023, Pittman2025}, this approximation is reasonable for our purposes. 

The luminosity of the circumplanetary disk is determined by viscous heating, where the viscosity is assumed to transport the mass that strikes the disk surface inward to $R_X$. 
Only a fraction $f_d$ of the mass accretion flow that enters the Hill sphere strikes the disk surface, with the rest hitting the planet directly. 
The total disk luminosity is then given by  
\begin{equation}\label{eq:Ld}
    L_d=f_d\frac{GM_p\dot{M}}{2R_X}\left(1-\frac{R_X}{R_C}\right)\,.
\end{equation}
The factor of $2$ in the denominator accounts for the rotation of the disk while the final term accounts for the outer boundary of the disk. 
This equation ignores the luminosity delivered to the disk by the material infalling from $R_H$ to $R_C$, assumes that the ballistic infall delivers all material to the disk outer radius, and assumes that the entire mass accretion rate onto the disk is transported inwards to accrete onto the planet. 
To conserve angular momentum, some negligible fraction of the material is instead transported outwards. 

At the inner edge of the disk, the magnetic fields deliver the accretion flow onto the (proto)planet. The planet is modeled as a single-temperature emitter with a luminosity equal to the total energy delivered to its surface by the accretion flow. The luminosity is then given by 
\begin{equation}\label{eq:Lp}
    L_p=\frac{GM_p\dot{M}}{R_p}\left(1-f_d\frac{R_p}{R_X}\right)\left(1-\frac{R_p^3}{3R_X^3}\right)\,.
\end{equation}
The second term in this equation accounts for the loss of energy to radiation and rotation in the disk and the third term accounts for the rotation of the planet itself. The effective surface temperature of the planet is then given by 
\begin{equation}\label{eq:Tp}
    T_p=\left(\frac{L_p}{4\pi\sigma R_p^2}\right)^{1/4}\,.
\end{equation}
This model implicitly assumes that the internal luminosity of the planet is negligible {and that the accretion luminosity is primarily released in the photosphere rather than in H$\alpha$ (see \citealt{Thanathibodee2019, Aoyama2021}) or the UV continuum (supported by \citealt{Finley2026})}.  

In addition to these generic results, there are two different models of the circumplanetary environment used in this work. The first, \RADp, is a two-dimensional numerical model that calculates the temperature and density structure of the circumplanetary disk. The second is a semi-analytic model (SAM) for the disk structure and radiative signatures. Throughout this work, \RADp will be treated as representing the ``true'' disk structure and radiative signatures and the SAM will be used to approximate these radiative signatures. 

\subsection{\RADp Model Structure}

\RADp is a two-dimensional iterative numerical model for the entire circumplanetary system. \RADp is based on the \RAD radiative transfer Monte Carlo model \citep{Dullemond2012}, with an additional iterative algorithm to ensure that the temperature and density of the circumplanetary disk are in both radiative and (vertical) hydrostatic equilibrium. This model also calculates the temperature of the circumplanetary envelope and the radiative signatures of the entire system. A detailed discussion of this algorithm can be found in Sec. 2 of \citet{Taylor2026}, which is summarized here.

In \RADp, the planet is modeled as a point-source emitter at the center of the system. The SED of the planet is found by interpolating the \texttt{SONORA Bobcat} models \citep{Marley2021} to the correct effective temperature $T_p$ (Eq. \ref{eq:Tp}) and surface gravity $g=GM_p/R_p^2$. The SED is downsampled to the model wavelengths. For wavelengths outside of the \qtyrange{0.5}{50}{\micro\meter} range provided by \texttt{SONORA Bobcat}, the SED is assumed to be a blackbody with the temperature $T_p$. 

The temperature and density structure of the disk is then calculated iteratively. For a given disk, the initial vertical density structure is taken from the D'Alessio Irradiated Accretion Disk model (\DIAD, \citealt{DAlessio1998, DAlessio1999, DAlessio2001, DAlessio2006, Micolta2024}). This density structure is loaded into \RAD and is heated by both the point-source planet at the system center and viscous heating throughout the disk. \RAD is then used to find the temperature of the disk, assuming thermal equilibrium. For a given temperature structure, requiring vertical hydrostatic equilibrium and viscous energy conservation fully specifies the disk density structure. By iteratively calculating the density and the temperature in turn, this model rapidly approaches a steady-state disk structure. 

The envelope is then added to the disk model by setting the density at every point in the Hill sphere to the maximum of the envelope and disk density at that point. The \RAD Monte Carlo algorithm is then used to calculate the temperature structure of the entire system and the emergent SEDs for a given viewing direction. 

\subsection{Semianalytic Model Structure}

Although \RADp is expected to be an accurate model of the structure and radiative signatures of these circumplanetary disks, the computational cost of a single model makes large-scale MCMCs prohibitive.\footnote{A single \RADp model takes approximately \qty{2}{hr} of wall time on a 32-core AMD Ryzen 9 9950X at \qty{5.7}{GHz}.} Although semianalytic models are much faster, previous thin-disk models do not reproduce the \RADp SEDs with sufficient accuracy. 

This paper constructs a semianalytic model (SAM) for the structure and radiative signatures of geometrically thick circumplanetary disks. When correctly calibrated, this model accurately reproduces \RADp SEDs while offering a significant speed increase (by a factor of \num{e8}). This section provides an abridged description of this model. A more detailed description, along with the accompanying mathematics, can be found in Appendix \ref{app:SAMmath}. 

In the SAM, the planet is simply modeled as a blackbody emitter, with a temperature given by Eq. \eqref{eq:Tp}. Given the low spectral resolution used in this work, the \texttt{SONORA Bobcat} SED used in \RADp is nearly identical to the relevant blackbody emitter.

% In addition to the system parameters, a given disk depends on three structural parameters --- $\dasp$, $q$, and $f_L$. These three parameters are critical in determining the SEDs of a given disk, but cannot be derived from the circumplanetary system parameters directly. Instead, they are interpolated across a grid of models. The models in this grid are generally referred to as the ``training models'', as they are used to calibrate the values of these structural parameters. 

The disk is assumed to locally emit as a blackbody. 
Between the inner radius $R_X$ and the outer radius $R_C$, the disk photosphere is set to follow a line of constant aspect ratio $\dasp=z_{\rm surf}/R$. 
The radiation from the disk is emitted from this surface, which is modeled as a plane-parallel slab. 
The surface temperature of the disk $T_d(R)$ is given by 
\begin{equation}
    T_d(R)=T_X\left(\frac{R}{R_X}\right)^{-q},
\end{equation}
where the parameter $q$ is the power-law slope of the temperature and $T_X$ is a normalization constant. Here, the value of $q$ is fixed to be between \num{0.75} (typical for a thin disk, \citealt{Shakura1973}) and \num{0.5} (characteristic of a flared disk), although some solar nebula models suggest $q\simeq1$ \citep{Lunine1982, Takata1996, Mosqueira2003a, Mosqueira2003b, Turner2014}. 

At the edges of the disk, the photosphere height is not zero. 
The boundaries of the disk define an inner wall at $R=R_X$ and an outer wall at $R=R_C$.
While the outer wall has a temperature $T_{\rm ow}=T_d(R_C)$, the inner wall is heated by the planet and has a temperature $T_{\rm iw}^4=T_d^4(R_X)+T_p^4(R_p/R_X)^2$.
This additional term accounts for the heating of the inner wall due to the planet's radiation. 
Due to its cold temperature, the emission of the outer wall is irrelevant (so long as $q>0.5$, which is forced to be true here) and ignored. 
The inner wall is modeled as an optically thick blackbody and is necessary for the SAM to accurately reproduce the SEDs at shallow viewing angles.

In order to ensure energy conservation, the temperature $T_X$ is chosen so that the total luminosity escaping from the disk in all directions is equal to the disk luminosity $L_d$. 
This constraint is parameterized by $\fL$, which is the ratio of the total luminosity escaping along the system pole $L(\psi=0)$ to the total system luminosity $L_p+L_d$. An iterative algorithm is then used to find the value of $T_X$ such that 
\begin{equation}
    L_p(\psi=0)+L_d(\psi=0)=\fL\cdot(L_p+L_d)\,.
\end{equation}
The parameter $\fL$ accounts for the self-shadowing of the system; correctly modeling this shadowing is necessary to calculate the correct SED of these thick disks. 

In addition to the system parameters given in Table~\ref{tab:canonvals}, the SAM requires these three additional ``structure parameters'' --- $\dasp$, $q$, and $\fL$ --- to fully define the disk temperature and radiative signatures. These parameters are not derived from first principles. Instead, these parameters are provided to the SAM from interpolations to a grid of \RADp models. This calibration ensures that the SAM disks are good matches to the \RADp disks. With these structure parameters, the disk's temperature is completely determined and the radiative signatures can be calculated at an arbitrary viewing angle $\psi$.

% In addition, the disk will be optically thick to the planet's radiation. 
% Since the planet emits isotropically, the disk will thus absorb some fraction of the planet's luminosity and reprocess it.
% The fraction of the planet's luminosity that is reemitted by the disk is characterized by the surface where the optical depth to the planet $\tau_p$ is equal to one. 
% Like the disk photosphere, the $\tau_p=1$ surface is also assumed to be given by a fixed aspect ratio.
% The aspect ratio $\plasp$ for the $\tau_p=1$ surface is generally larger than the photosphere aspect ratio $\dasp$.

The geometrically thick disk also shadows part of the envelope from the planet's radiation and, depending on the viewing angle, obscure part of the envelope from the observer. The envelope can thus be divided into three regions. The geometry of this system is shown in panel b of Fig.~\ref{fig:cpdgeom}. Region 1 consists of the cone of material at the pole of the system interior to the disk, which can only be seen if the observer is close to the system pole. Region 2 is the truncated cone between the disk faces, which is heated directly by the planet and is visible to the observer at all positions. The remainder of the envelope, which is shadowed by the disk, is assigned to Region 3. 

The temperatures of each of these three regions are treated differently. Regions 1 and 2 are heated directly by the planet and have a spherically symmetric temperature distribution. In particular, their temperature distribution takes the form \citep{Adams1985}
\begin{equation}
    T_e(r)=T_C\left(\frac{r}{R_C}\right)^{-2/(4+\gamma)}\,
\end{equation}
where $\gamma=1/2$ is the power-law slope of the opacity used here. This equation applies if the envelope is optically thin to its own radiation, which has been shown to be true so long as $\dot{M}\lesssim\qty{30}{M_J/Myr}$ \citep{Taylor2025}. In contrast, the temperature of Region 3 is set to be $T_d(R_C)$, since it is significantly shadowed by the disk. 

The observability of the regions is also different. The envelope is assumed to be optically thin to its own radiation, so that the emergent SEDs can be calculated by simply integrating over the envelope. Region 1 is only visible if the surface of the disk does not interfere (so that the viewing angle $\psi<\arctan(1/\dasp)$), while Regions 2 and 3 are assumed to always be visible to an observer. 

The only undefined parameter is the temperature normalization $T_C$, which is also specified by energy conservation. The envelope absorbs radiation from the planet and the disk and reprocess it to longer wavelengths. The total energy absorbed by the envelope is calculated and then $T_C$ is chosen so that the viewing-angle-integrated energy is equivalent. Once again, the algorithms for calculating $T_C$ and the observed SEDs are detailed in Appendix \ref{app:SAMmath}.

\begin{figure}
    \centering
    \includegraphics{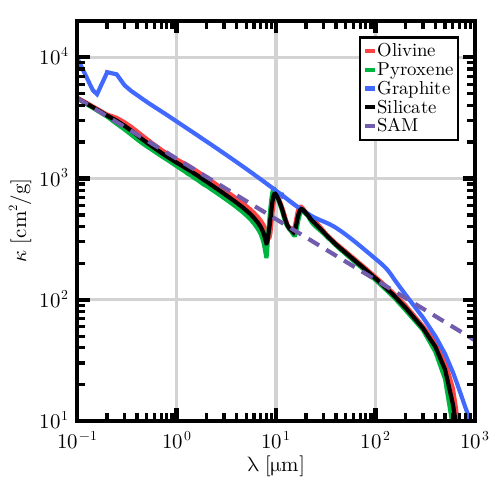}
    \vspace{-10pt}
    \caption{\textbf{Opacity Comparison.} A comparison of the wavelength-dependent dust opacity used in \RADp and the SAM. The \RADp dust opacity is composed of a weighted average of silicates (olivine and pyroxene, red and green respectively, with the combination shown in black) and graphite (blue). The SAM opacity is assumed to be a power law with a slope of $\kappa\propto\lambda^{-0.5}$ (purple, dashed). Except for the \qty{10}{\micro\meter} silicate feature, this power-law approximation is generally appropriate for $\lambda\lesssim\qty{300}{\micro\meter}$. At longer wavelengths, the power-law opacity significantly overestimates the opacity of the dust. This figure shows the opacity in terms of per unit mass of dust.}
    \vspace{-10pt}
    \label{fig:opas}
\end{figure}

\subsection{Model Opacities}

It is worthwhile to briefly discuss the opacities used in these models. The opacity of dust in the circumplanetary environment is highly uncertain and has a significant impact on the system radiative signatures. 

In \RADp, the dust is modeled as a single population with a power-law size distribution $n(a)\propto a^{-3.5}$, a minimum size of \qty{0.005}{\micro\meter}, and a maximum size of \qty{100}{\micro\meter} (e.g., \citealt{Mathis1977}). The scattering opacity is directly added to the absorptive opacity to create a total effective opacity. The opacity is taken from \citet{DAlessio2001} and \citet{Micolta2024}. The dust is characterized by the dust-to-gas mass ratio $\eta$ and the silicate fraction $f_{\rm Si}$. The total opacity per unit dust mass is found by combining the silicate component (\qty{50}{\percent} each of olivine and pyroxene) with a graphite component, weighted by the silicate fraction. Multiplying by $\eta$ gives the total effective opacity per unit disk mass. These opacities are shown in Fig. \ref{fig:opas}. 

In the SAM, the opacity is modeled as a power law, so that 
\begin{equation}
    \kappa(\nu)=\kappa_0\left(\frac{\nu}{\nu_0}\right)^\gamma\,.
\end{equation}
For $\nu_0=\qty{3e13}{Hz}$, the normalization is $\kappa_0=\qty{462}{cm^2/g}$ for the silicates and $\kappa_0=\qty{923}{cm^2/g}$ for graphite. These values represent the cross-section per gram of dust, with the per-gas opacity given by $\eta\kappa$. These opacities are again combined using $f_{\rm Si}$. Here, the power-law slope is set to $\gamma=0.5$. This opacity is also shown in Fig. \ref{fig:opas} (for $f_{\rm Si}=1$).

The SAM opacity matches the \RADp opacity relatively well over most of the domain. If the silicate fraction is large, the power-law opacity cannot reproduce the \qty{10}{\micro\meter} silicate feature, but is otherwise generally accurate for $\lambda\leq\qty{300}{\micro\meter}$. At longer wavelengths, the opacity of both silicate and graphite dust drops off quickly, while the power-law opacity continues. These wavelengths are generally not important, but this effect is noticeable in some regimes. 

\begin{figure*}[t!]
    \centering
    \includegraphics{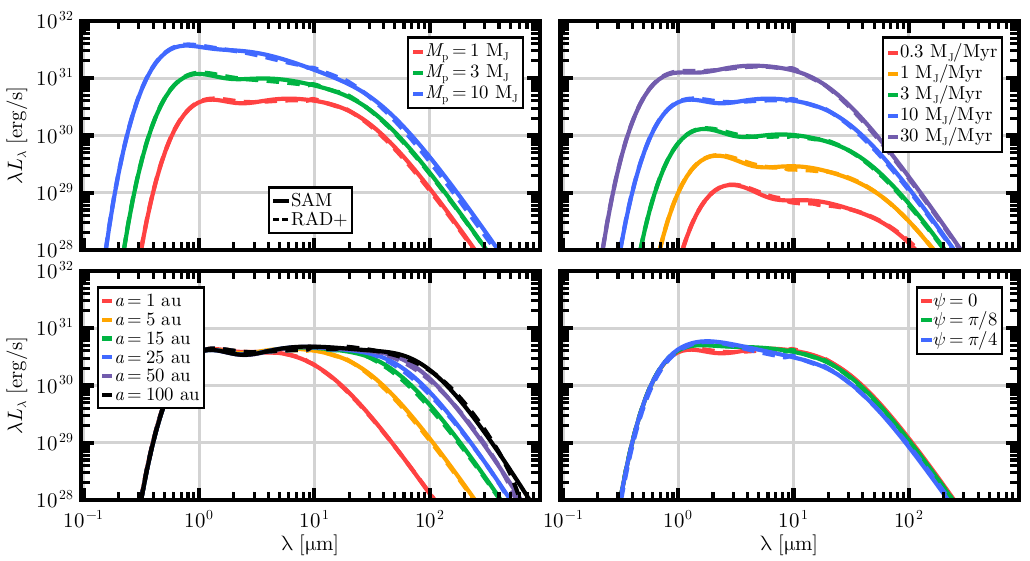}
    \vspace{-10pt}
    \caption{\textbf{Model SEDs Comparison.} A comparison of the SEDs calculated for a geometrically thick disk (solid) and the numerical \RADp model (dashed). The effects of planet mass, mass accretion rate, orbital semimajor axis, and viewing angle are shown in different panels. Unless otherwise specified, all system parameters are set to their fiducial values (Table \ref{tab:canonvals}). All SEDs shown do not include the attenuation of the circumplanetary envelope. }
    \vspace{-10pt}
    \label{fig:specdiskcompgrid}
\end{figure*}

\subsection{Comparing Semianalytic and Numerical Models}

Compared to previous thin-disk semianalytic models, the model introduced in this work more accurately reproduces the numerical SEDs predicted by the \RADp model. Fig.~\ref{fig:specdiskcompgrid} compares the SEDs predicted by the thick-disk SAM and the ``true'' \RADp model. Across parameter space, the SAM matches the \RADp SEDs reasonably well.

The \RADp and SAM models disagree in the limit of large $a$ and $\lambda\gtrsim\qty{300}{\micro\meter}$. In this region of parameter space, the actual opacity of the dust drops off sharply compared to the power law assumed in the SAM (see Fig.~\ref{fig:opas}). As the opacity assumption breaks down, the \RADp disk becomes optically thin and the emitted radiation is significantly reduced. The unconstrained opacity dependence means that fitting observations in this wavelength range with the SAM will be difficult.

The values of $\dasp$, $q$, and $\fL$ are identical between the SAM and \RADp models shown in this figure. For a random point in parameter space, the values of these parameters used in the SAM will be interpolated across parameter space and may be inaccurate. Although the close agreement of these SEDs indicates that the structure of the SAM is correct, this figure does not imply anything about the accuracy of the interpolation of these structure parameters. 

\begin{figure*}[t!]
    \centering
    \includegraphics{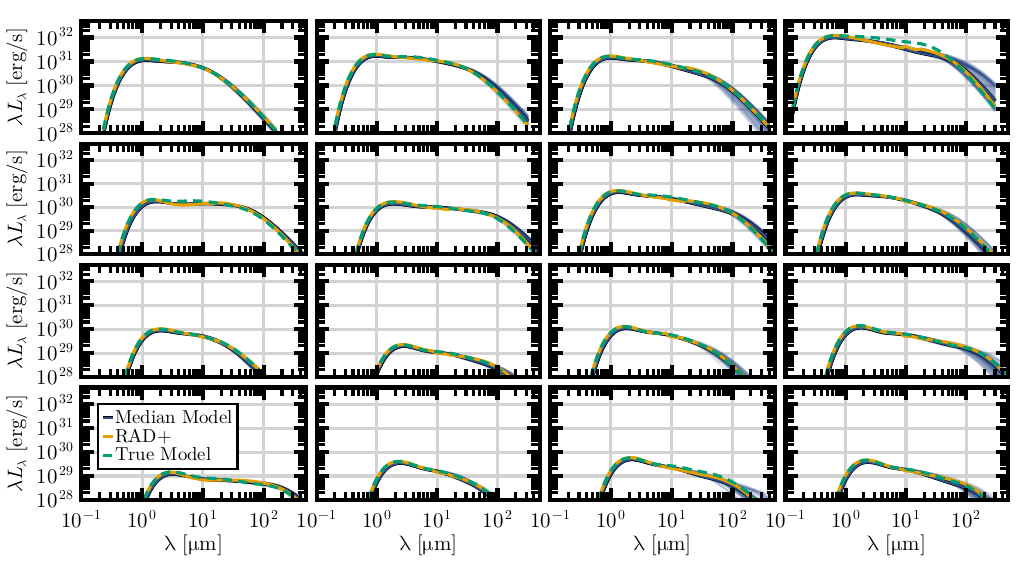}
    \vspace{-10pt}
    \caption{\textbf{SED Comparison.} Comparing the synthetic \RADp (orange), true-parameter SAM (green), and MCMC (blue) SEDs for a grid of models. The model parameters are all randomized, but the mass accretion rate increases in the vertical direction and the planet mass increases horizontally. In general, these SEDs are in good agreement. Even in cases where the SEDs disagree slightly, the optimal parameters are within \qty{1}{\sigma} of the true values.}
    \vspace{-10pt}
    \label{fig:specgrid}
\end{figure*}

To test this interpolation, we use \RADp to generate the SEDs of 40 ``test models'', which are randomly distributed in parameter space and observed from random angles. The structural parameter values used in the test models are interpolated from a grid of 40 \RADp ``training models'', which cover all of parameter space. Fig.~\ref{fig:specgrid} compares the \RADp and SAM SEDs for 16 of these test models, showing the accuracy of the interpolation. 

The test models also provide an estimate of the inherent uncertainty in the model, which can be characterized as the average root-mean-square error between the full \RADp and SAM SEDs evaluated at the correct locations in parameter space. Across these 40 ``test models'' and for $\qty{0.1}{\micro\meter}\leq\lambda\leq\qty{e3}{\micro\meter}$, the model error is approximately \qty{20}{\percent}. This result demonstrates that the interpolation of the structural parameters has little impact on the agreement between the SAM and \RADp models. This inherent uncertainty will also be accounted for in the future MCMC fits using this model.

\begin{table}[t!]
    \centering
    \caption{\textbf{Fiducial Model Parameters.} The model parameters of the three fiducial models explored in detail in this work. The ``young Jupiter'' model has $M_p\simeq\qty{1}{M_J}$, $\dot{M}\simeq\qty{3}{M_J/Myr}$, and $a\sim\qty{5}{au}$. The ``young super-Jupiter'' model has $M_p\simeq\qty{1}{M_J}$, $\dot{M}\sim\qty{10}{M_J/Myr}$, and $a\sim\qty{20}{au}$. Finally, the ``mature super-Jupiter'' model has $M_p\sim\qty{10}{M_J}$, $\dot{M}\sim\qty{0.1}{M_J/Myr}$, and $a\sim\qty{50}{au}$. All three systems are observed with a viewing angle of $\psi=\qty{30}{\degree}$. All other model parameters are either set to their default values ($M_\star$, $R_p$, $B_{p,0}$, $\lambda_d$) or randomly chosen from the domain ($\alpha$, $\eta$, $f_{\rm Si}$). }
    \begin{tabular}{c|D{.}{.}{3.2}D{.}{.}{3.2}D{.}{.}{3.2}}
        % \multirow{2}{*}{Parameter} & \multirow{2}{=}{\makecell{Young \\ Jupiter}} & \multirow{2}{=}{\makecell{Young Super-\\ Jupiter}} & \multirow{2}{}{\makecell{Mature Super-\\ Jupiter}}\\
        % \phantom{-} & & & \\ \hline
        % \multirow{2}{*}{Parameter} & \multicolumn{1}{p{2cm}}{Young Jupiter} & \multirow{2}{*}{\makecell{Young Super-\\ Jupiter}} & \multirow{2}{*}{\makecell{Mature Super-\\ Jupiter}}\\
        \multirow{2}{*}{Parameter} & \multicolumn{1}{c}{Young} & \multicolumn{1}{c}{Young Super-} & \multicolumn{1}{c}{Mature Super-} \\[-2pt]
        \phantom{-} & \multicolumn{1}{c}{Jupiter} & \multicolumn{1}{c}{Jupiter} & \multicolumn{1}{c}{Jupiter} \\\hline
        $M_p$ [\unit{M_J}] & 0.96 & 1.04 & 8.78 \\
        $\dot{M}$ [\unit{M_J/Myr}] & 3.07 & 7.65 & 0.37 \\
        $a$ [\unit{au}] & 6.30 & 17.66 & 62.93 \\
        $\psi$ [\unit{\degree}] & 30.00 & 30.00 & 30.00\\
        $M_\star$ [\unit{M_\odot}] & 1.00 & 1.00 & 1.00 \\
        $R_p$ [\unit{R_J}] & 1.4 & 1.4 & 1.4 \\
        $\alpha$ & 0.080 & 0.098 & 0.049 \\
        $\eta$ & 0.0100 & 0.0098 & 0.0090 \\
        $f_{\rm Si}$ & 0.036 & 0.300 & 0.780 \\
        $B_{p,0}$ [\unit{G}] & 500 & 500 & 500 \\
        $\lambda_d$ & 1.0 & 1.0 & 1.0 \\
    \end{tabular}
    \label{tab:exemplarmodels}
\end{table}

\section{Fits to Synthetic Observations}\label{sec:synthobv}

A critical question is how well observations can constrain the properties of forming giant planets. To this end, an MCMC algorithm is used to compare the thick-disk semianalytic model to synthetic observational SEDs generated from \RADp. We assume that the stellar mass, orbital semimajor axis, viewing angle, and protoplanet radius are known \textit{a priori}. First, the stellar mass can be measured independently of observations of the planet. The orbital semimajor axis can be derived by both astrometry and projected-distance calculations and the viewing angle can be found by assuming that the planet's spin pole is normal to the disk. {Note that these models also make specific assumptions regarding the dust size distribution and settling, which will have some impact on the resulting SEDs.} The model parameters $M_p$, $\dot{M}$, $B_{p,0}$, $f_{\rm Si}$, $\eta$, and $\alpha$ are fit to the observations. The MCMCs are run on \num{30} walkers for \num{10000} steps with a \num{1000} step burn-in and uniform priors. 

There are several parameters that are not varied in this model, which may affect the SEDs. The first parameter is the planet radius $R_p$. Although most young giant planets are expected to have $R_p\simeq\qty{2}{R_J}$ (e.g., \citealt{Batygin2025, Knierim2026}), different radii can change the planet temperature and system luminosity for a given mass and mass accretion rate. These changes may have consequences for the SEDs and increase the uncertainty of model fits. The second parameter is the outer radius of the disk $R_C$, which is here set to the centrifugal radius. If the infalling material has an angular momentum deficit $\lambda_d$, the outer edge of the disk has a radius given by $\lambda_d^2 R_C$ \citep{Adams2025}. Of course, introducing either $\lambda_d$ or $R_p$ as another free parameter would necessarily increase the uncertainty in the model and weaken the constraints on the other parameters. 

The increase in uncertainty from including the planet radius and the angular momentum deficit is moderate at worst. After all, the variations in $R_p$ and $\lambda_d^2$ are relatively small. The planet radius $R_p$ certainly cannot be less than \qty{1}{R_J} and is probably not larger than \qty{3}{R_J}, which is only a difference of $\sim\qty{0.5}{dex}$. If $R_p$ were included as a free parameter in the MCMC runs, then $L_{\rm tot}$ would still be constrained to the same magnitude, but the product $M_p\dot{M}$ would have an additional minor scatter of \qty{0.25}{dex} on average. On the other hand, $\lambda_d$ is can be as low as \num{0.3} and cannot be larger than \num{1.0}. The variations in the outer disk radius can be at most \qty{1}{dex}, with an expected variation of \qty{0.5}{dex}. While an order of magnitude difference can be significant, the domains of $M_p$ and $\dot{M}$ span orders of magnitude. The variation in these parameters would therefore likely dominate over the comparatively small impacts of $R_p$ and $\lambda_d$. 

For each of the test models, the SED of the posterior is a close match to the \RADp SEDs. A comparison of the MCMC results, the true-parameter model SED, and the \RADp SED for 16 of the test models are shown in Fig.~\ref{fig:specgrid}. This figure is sorted so that high-mass planets are at the top and high mass accretion rates on the right. First, this figure confirms that the SAM is relatively accurate at matching the \RADp SEDs, even when the structure parameters are calculated via interpolation. Second, the SEDs from the MCMC posterior are in good agreement with the \RADp SEDs. While not shown, the parameter posteriors are also all within \qty{1}{\sigma} from the true value.

There are occasional systematic errors in the SEDs, particularly in the MIR near \qty{10}{\micro\meter}. This region is where the power-law opacity used in the SAM is a poor match for the \RADp dust opacity, which includes the \qty{10}{\micro\meter} silicate feature. When there are systematic differences between the \RADp and SAM SEDs, the most common issue is that the SAM SEDs are slightly too bright in this wavelength region. This difference is the result of a slightly higher (and colder) photosphere than anticipated due to the dust opacity spike. These disagreements are difficult to fix without a thorough understanding of the two-dimensional temperature structure of the disk, which is incompatible with a simple semianalytic model. 

\begin{figure}
    \centering
    \includegraphics{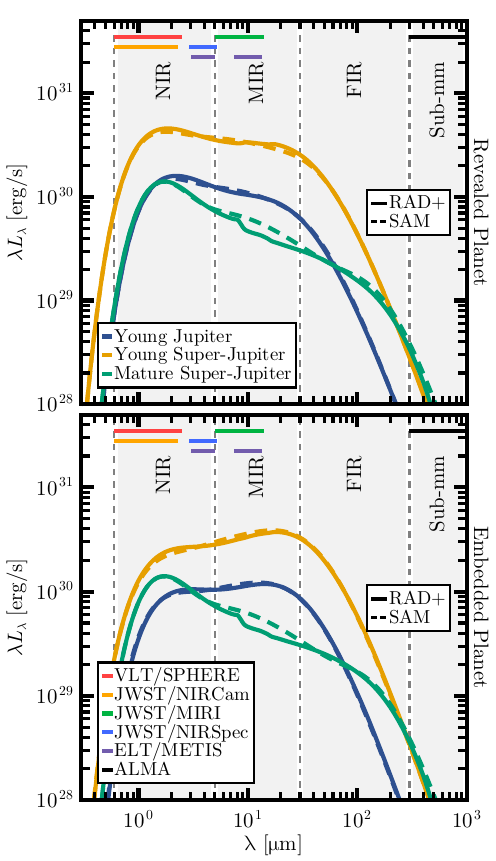}
    \vspace{-10pt}
    \caption{\textbf{Model SED Comparison.} The SEDs as calculated by \RADp (solid) and the SAM (dashed) for the three fiducial models. The top panel shows the SEDs when there is no circumplanetary envelope (e.g., when the planet is in a deep gap or accreting via a streamer). The bottom panel shows the SEDs when the envelope is included and the planet is embedded in the circumstellar disk. The \RADp and SAM SEDs generally agree reasonably well. The wavelength bands under consideration (NIR, MIR, FIR, and sub-mm) are also shown. The colored lines on the top show the wavelength coverage of several important instruments (VLT/SPHERE, JWST/NIRCam, JWST/MIRI, JWST/NIRSpec, ELT/METIS, and ALMA). }
    \vspace{-10pt}
    \label{fig:exempmodelcomp}
\end{figure}

Although the SAM works across essentially all of the relevant parameter space, it is not feasible to discuss each of the forty test models in detail. For practical purposes, we focus on three models which represent points of interest in parameter space. These exemplars are (i) a young Jupiter, (ii) a young super-Jupiter, and (iii) a mature super-Jupiter. The young Jupiter system is set to have a mass slightly less than \qty{1}{M_J}, a mass accretion rate of approximately \qty{3}{M_J/Myr}, and an orbit at \qty{6.3}{au}. The young super-Jupiter has a mass of \qty{1}{M_J} and a much higher accretion rate of approximately \qty{8}{M_J/Myr}. It orbits at $a\simeq\qty{20}{au}$, where it may be possible to observe this system. Finally, the mature super-Jupiter has a mass of \qty{9}{M_J}, a low mass accretion rate of $\dot{M}=\qty{0.4}{M_J/Myr}$, and an orbit at \qty{63}{au}. The parameters of these models are summarized in Table~\ref{tab:exemplarmodels} and the SEDs calculated by \RADp and the SAM are shown in Fig.~\ref{fig:exempmodelcomp}. 

% It is also worthwhile to consider the constraints that can be obtained from observations in different wavelength bands. Four different wavelength bands are considered --- the full SED (\qtyrange{0.1}{300}{\micro\meter}), the near-infrared (NIR, \qtyrange{0.6}{5}{\micro\meter}), mid-infrared (MIR, \qtyrange{5}{30}{\micro\meter}), and far-infrared (FIR, \qtyrange{30}{300}{\micro\meter}). Current instruments exist to observe in the NIR and MIR, but FIR capability has been lost since the ends of the Herschel and SOFIA missions. While sub-millimeter observations at $\lambda\geq\qty{300}{\micro\meter}$ are within the capabilities of several instruments, the large differences in model opacity at these wavelengths (see Fig. \ref{fig:opas}) and the uncertainty of the opacity means that this wavelength range is unsuitable for fitting. 

The SAM generally accurately reproduces the \RADp SEDs for the example models. The one exception is the mature super-Jupiter, where the \RADp SED is slightly dim near \qty{10}{\micro\meter}. This is a manifestation of the silicate opacity feature at this wavelength and does not have a significant impact on these results. This figure also shows several critical wavelength bands under consideration --- the near-infrared (NIR, \qtyrange{0.6}{5}{\micro\meter}), mid-infrared (MIR, \qtyrange{5}{30}{\micro\meter}), and far-infrared (FIR, \qtyrange{30}{300}{\micro\meter}). Current and confirmed instruments observe in the NIR (VLT/SPHERE, JWST/NIRCam, JWST/NIRSpec, and ELT/METIS) and the MIR (JWST/MIRI, ELT/METIS), but FIR capability has been lost since the ends of the Herschel and SOFIA missions. Although sub-millimeter observations at $\lambda\geq\qty{300}{\micro\meter}$ are within the capabilities of several instruments (e.g. ALMA), the large differences in model opacity at these wavelengths (see Fig. \ref{fig:opas}) and the uncertainty of the opacity normalization means that this wavelength range is unsuitable for fitting. 

The NIR is generally exclusively sensitive to the short-wavelength emission from the planet, the MIR is sensitive to the hot inner circumplanetary disk, and the FIR is sensitive to the outer edges of the disk. The sub-mm is sensitive almost entirely to the optically thin outermost edges of the disk where the dust opacity becomes important. 

For each planet, there are two possible accretion configurations that can be considered. In the first case, the planet resides in a gap in the circumstellar disk and is accreting via a stream of material. Since the circumplanetary envelope is a direct result of the diffuse infalling material, a streamer-fed system is not be embedded in a circumplanetary envelope. The SEDs of such a system only have the components due to the planet and the disk, which would not be attenuated by the circumplanetary envelope. Instead, this planet is revealed to observations from above or below the disk plane. 

Although it will be substantially more difficult, it may also be possible to observe a forming planetary system while it is embedded in the natal circumstellar disk. For such a system, the diffuse accretion flow generates the circumplanetary envelope which attenuates the planet and disk. For each model, the SAM is also fit to synthetic observations in these circumstances. We assume that the further attenuation due to the circumstellar disk has been corrected, so that the SEDs are the emergent luminosity from the Hill sphere. 

For each example model, wavelength band, and accretion configuration, we generate synthetic observations and use an MCMC to determine the constraints on observed parameters. When running the MCMC fits, all SEDs are downsampled to \num{400} wavelength points (representing $\Delta\lambda/\lambda\simeq300$ in the NIR). This ensures that the amount of data available has no impact on the constraints and only the informational content of the wavelength band is considered. 

The following sections explore how observations in each of these wavelength bands can be used to constrain the properties of forming planetary systems and how these constraints depend on the accretion configuration.

\begin{figure*}[t!]
    \centering
    \vspace{-30pt}
    \includegraphics{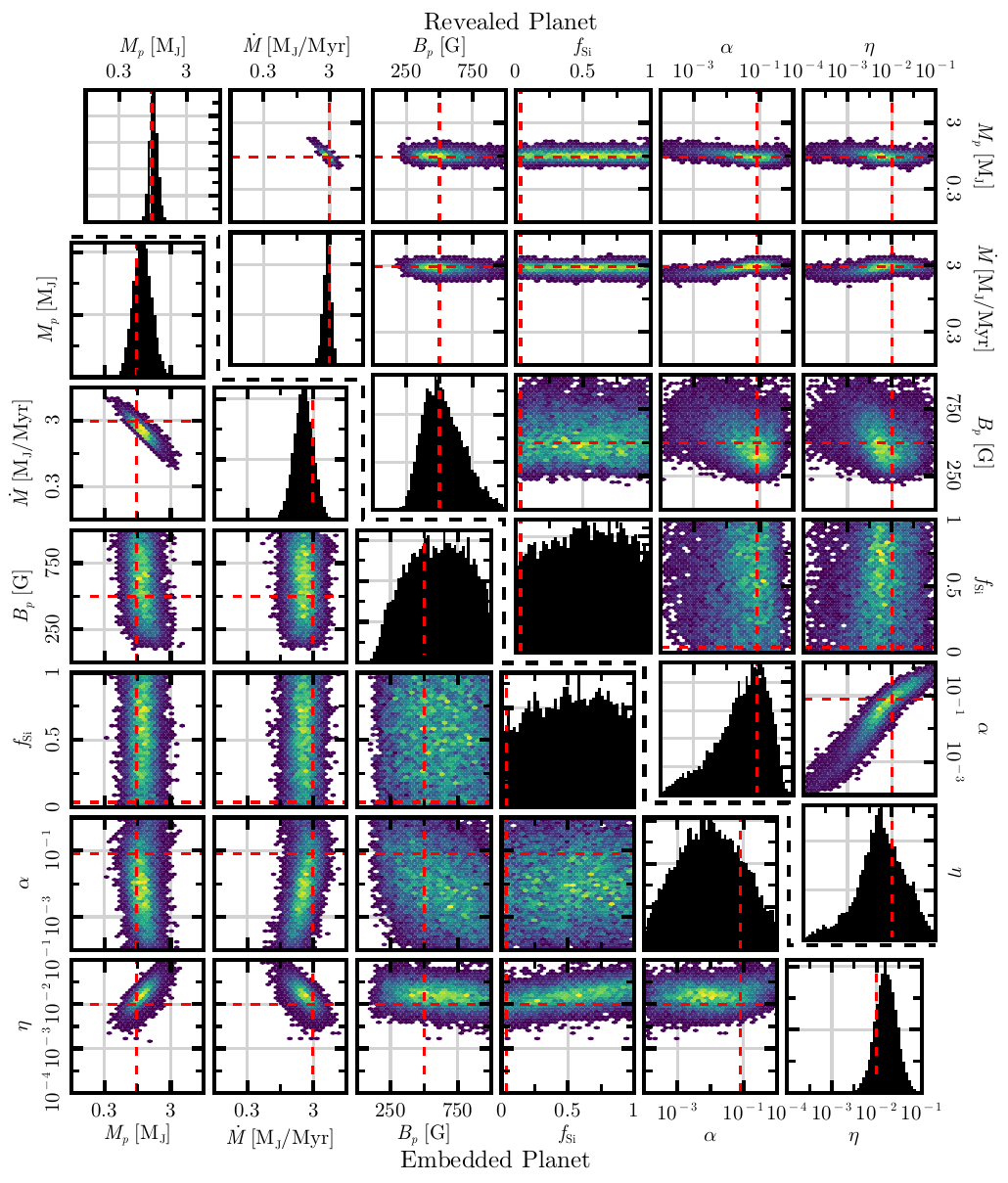}
    \vspace{-20pt}
    \caption{\textbf{Cornerplots of the Full SED.} The cornerplots resulting from using an MCMC to fit the SAM to the entire synthetic SED of a young Jupiter and its circumplanetary disk. The upper right corner shows the cornerplot for a revealed planet and the bottom left corner shows the cornerplot for an embedded planet. The colors show the probability distribution and the red lines show the true values of the parameters. All parameters that are not shown are set to their true values. In general, the MCMC has difficulty constraining any individual parameter, although it is able to constrain $M_p\dot{M}$ reasonably well. }
    \vspace{-15pt}
    \label{fig:YJcorner}
\end{figure*}

\begin{figure*}[t!]
    \centering
    \vspace{-30pt}
    \includegraphics{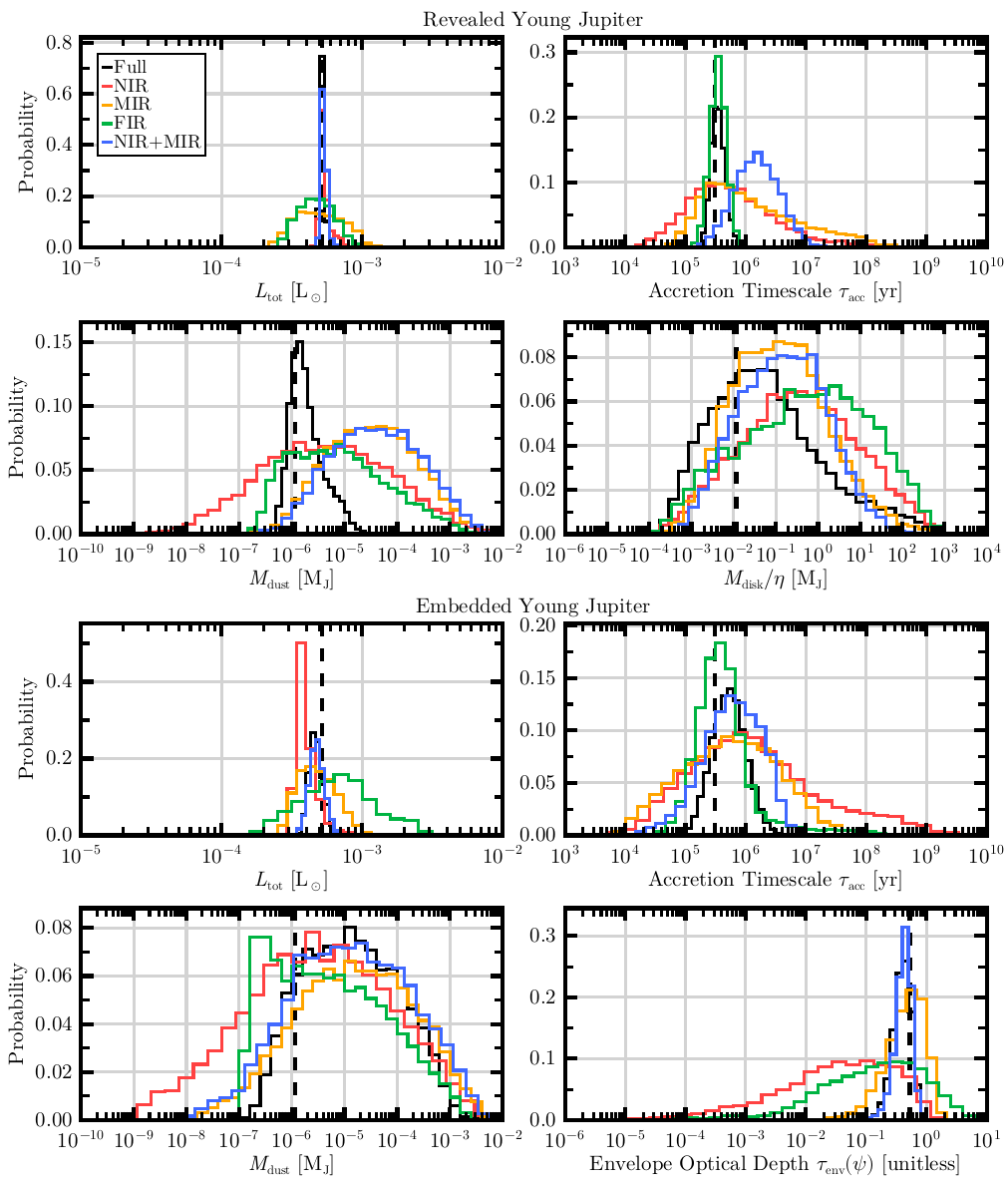}
    \vspace{-5pt}
    \caption{\textbf{Young Jupiter Derived Parameters.} Posteriors of several important parameters for observations of a revealed and an embedded young Jupiter over a range of wavelengths. The total luminosity $L_{\rm tot}$ and the accretion timescale $\tau_{\rm acc}$ are linearly independent and jointly define $M_p$ and $\dot{M}$, while $M_{\rm dust}$ and $M_{\rm disk}/\eta$ jointly define $\alpha$ and $\eta$. For a revealed system, observations are capable of putting tighter constraints on $L_{\rm tot}$ and $M_{\rm dust}$ than on $\tau_{\rm acc}$ or $M_{\rm disk}/\eta$. For an embedded system, the dust mass is poorly constrained and the envelope optical depth $\tau_{\rm env}$ is reasonably well-constrained. The colors show the fits for observations in different wavelength bands---the NIR (red), MIR (orange), FIR (green), combined NIR and MIR (blue), and the full SED (black). }
    \vspace{-15pt}
    \label{fig:YJhists}
\end{figure*}

\subsection{Young Jupiter}

We first consider the full ($\lambda=\qtyrange{0.6}{300}{\micro\meter}$) SED of a young Jupiter. Cornerplots of the results for both a revealed and embedded planet are shown in  Fig.~\ref{fig:YJcorner}.
% The upper right corner of this figure shows the cornerplot for a revealed system, while the lower left is the embedded cornerplot.

Although the MCMC posteriors agree with the true parameters to within \qty{1}{\sigma}, the constraints on model parameters are somewhat loose. For a revealed young Jupiter, the best-constrained parameters (the planet mass $M_p$ and mass accretion rate $\dot{M}$) are only constrained to approximately \qty{1}{dex}. The magnetic field strength $B_{p,0}$, silicate fraction $f_{\rm Si}$, viscous parameter $\alpha$, and dust-to-gas ratio $\eta$ are barely constrained over their domain. 

Although the parameters cannot be constrained separately, the revealed system SEDs provide significant joint constraints on two parameter pairs, specifically $M_p$/$\dot{M}$ and $\alpha$/$\eta$. In both cases, one derived parameter is constrained tightly while a linearly independent parameter is poorly constrained. For $M_p$ and $\dot{M}$, the posterior generally constrains $M_p\dot{M}$ (which is proportional to the total system luminosity $L_{\rm tot}$, see also \citealt{Shibaike2024}) and only loosely constrains the accretion timescale {$\tau_{\rm acc}=M_p/\dot{M}$}. On the other hand, the well-constrained $\eta/\alpha$ is proportional to the total disk dust mass $M_{\rm dust}$, since for an $\alpha$-disk $\alpha\propto 1/\Sigma\propto M_{\rm disk}$. The orthogonal parameter $1/(\eta\,\alpha)$ is proportional to $M_{\rm disk}/\eta$, which has no additional meaning. 

For an embedded system, the constraints are generally qualitatively similar to that of a revealed system. In general, observations including an envelope --- where the system is embedded in the circumstellar disk --- provide less stringent constraints than observations of a system embedded in a disk gap. The envelope reprocessing removes information from the system SED by allowing for absorption to modify the spectral slopes. The inclusion of the envelope thereby increases degeneracy and loosens the parameter constraints.

The one exception is the dust-to-gas ratio. For an embedded system, the previously-seen correlation between $\alpha$ and $\eta$ is broken. Instead, $\eta$ is constrained to within \qty{1}{dex} and there are weak correlations between $M_p$, $\dot{M}$, and $\eta$. These three parameters are proxies for the optical depth of the envelope to the planet, which is defined to be 
\begin{equation}
    \tau_{\rm env}(\psi)=\int\displaylimits_{R_p}^{R_H}\!\!\rho(r,\psi)\kappa_P(T_p)\,\de r\,.
\end{equation}
Since the dust is the dominant source of opacity, the total opacity scales directly with $\eta$. At the same time, the envelope density $\rho\propto\dot{M} M_p^{-1/2}$ (see  \citealt{Ulrich1976} etc.), so that the optical depth depends on these three parameters. Therefore, observations of these systems may be able to constrain the optical depth of the envelope. The constraints on these derived parameters ($L_{\rm tot}$, $\tau_{\rm acc}$, $M_{\rm dust}$, and $M_{\rm disk}/\eta$ or $\tau_{\rm env}$) for both revealed and embedded systems are shown in Fig.~\ref{fig:YJhists}. 

Additional wavelength information more tightly constrains all relevant parameters. As expected from Fig.~\ref{fig:YJcorner}, the SEDs provide a significantly stronger constraint on the total luminosity than they do on the accretion timescale. For the full SED of a revealed planet, the luminosity is \qty{3}{\sigma} constrained to $\Delta L_{\rm tot}\simeq\qty{0.045}{dex}$ and $\Delta\tau_{\rm acc}\simeq\qty{0.35}{dex}$. In parallel, the dust mass $M_{\rm dust}$ is constrained to \qty{1}{dex} and $M_{\rm disk}/\eta$ is only constrained to within \qty{4}{dex}. For an embedded planet, the constraints on $L_{\rm tot}$ and $\tau_{\rm acc}$ attain roughly the same accuracy, although the constraints are slightly less precise ($\Delta L_{\rm tot}\simeq\qty{0.15}{dex}$ and $\Delta\tau_{\rm acc}\simeq\qty{0.85}{dex}$). 

This figure also shows that different wavelength bands provide different information concerning the system parameters. The first panel in Fig.~\ref{fig:YJhists} shows that the NIR contains most of the luminosity information for a revealed planet. Of course, since the NIR captures the planet SED and the planet emits most of the system luminosity, this behavior is expected. The dust mass is approximately equally well-constrained by all three wavelength bands, with only the full SED providing the optimal constraint. Since the dust emission is spread out over all three bands, only broad wavelength coverage can effectively constrain the disk dust mass. A similar principle applies to $M_{\rm disk}/\eta$, although even the full SED provides only a weak constraint. 

For the embedded planet, joint NIR and MIR SEDs provide the strongest constraints on $L_{\rm tot}$, in contrast to the revealed system where the NIR dominates. When the envelope is included, it absorbs much of the NIR radiation and re-emits it in the MIR, flattening the SED. As a result, the NIR does not contain a majority of the emission and the NIR and MIR combined are necessary to capture most of the system luminosity. 

The constraints on the dust mass $M_{\rm dust}$ are weaker for an embedded young Jupiter than a revealed system, even when the entire SED is available. The envelope absorption and emission reprocesses the SEDs and somewhat obscures the total dust mass. Instead, these observations put a \qty{0.45}{dex} constraint on the line-of-sight optical depth to the planet $\tau_{\rm env}$. For individual wavelength bands, the MIR provides most of the information on $\tau_{\rm env}$, since the envelope generally radiates in the MIR. However, the envelope emission in the MIR could be accounted for by a bright disk. Joint observations in the NIR break the minor degeneracy between the disk and the envelope, since the disk is necessarily colder than the planet. The spectral slope across the NIR and MIR thus provides a strong diagnostic for envelope absorption and re-emission. 

In parallel, the FIR, which is sensitive primarily to the outer disk, provides the tightest constraints on $\tau_{\rm acc}$ for both revealed and embedded planets. The radiative signatures in this band are sensitive to both the radius and temperature of the outer edge of the disk, which depend on $M_p$ and $\dot{M}$ separately. Specifically, for a fixed $a$ and $M_\star$, $R_C$ depends exclusively on $M_p$ while the temperature at the outer edge depends on the temperature power-law index $q$. The value of $q$ further depends on the geometric thickness of the disk, since a geometrically thin disk has $q\simeq0.75$ and a geometrically thick disk has $q\simeq0.5$. The geometrical thickness of the disk, and thus the value of $q$, depends primarily on $\dot{M}/M_p$. For increased $\dot{M}$, the disk must be hotter --- and geometrically thicker --- to account for the increased viscous accretion and luminosity. For larger $M_p$ the gravitational force towards the midplane is stronger and the disk is geometrically thinner. The dependence of the outer disk SEDs on $R_C$ and $q$, both of which depend on $M_p$ and $\dot{M}$ separately, enables the FIR to provide $\qty{0.4}{dex}$ (revealed) or $\qty{1.6}{dex}$ (embedded) constraints on the accretion timescale.

\begin{figure*}
    \centering
    \vspace{-30pt}
    \includegraphics{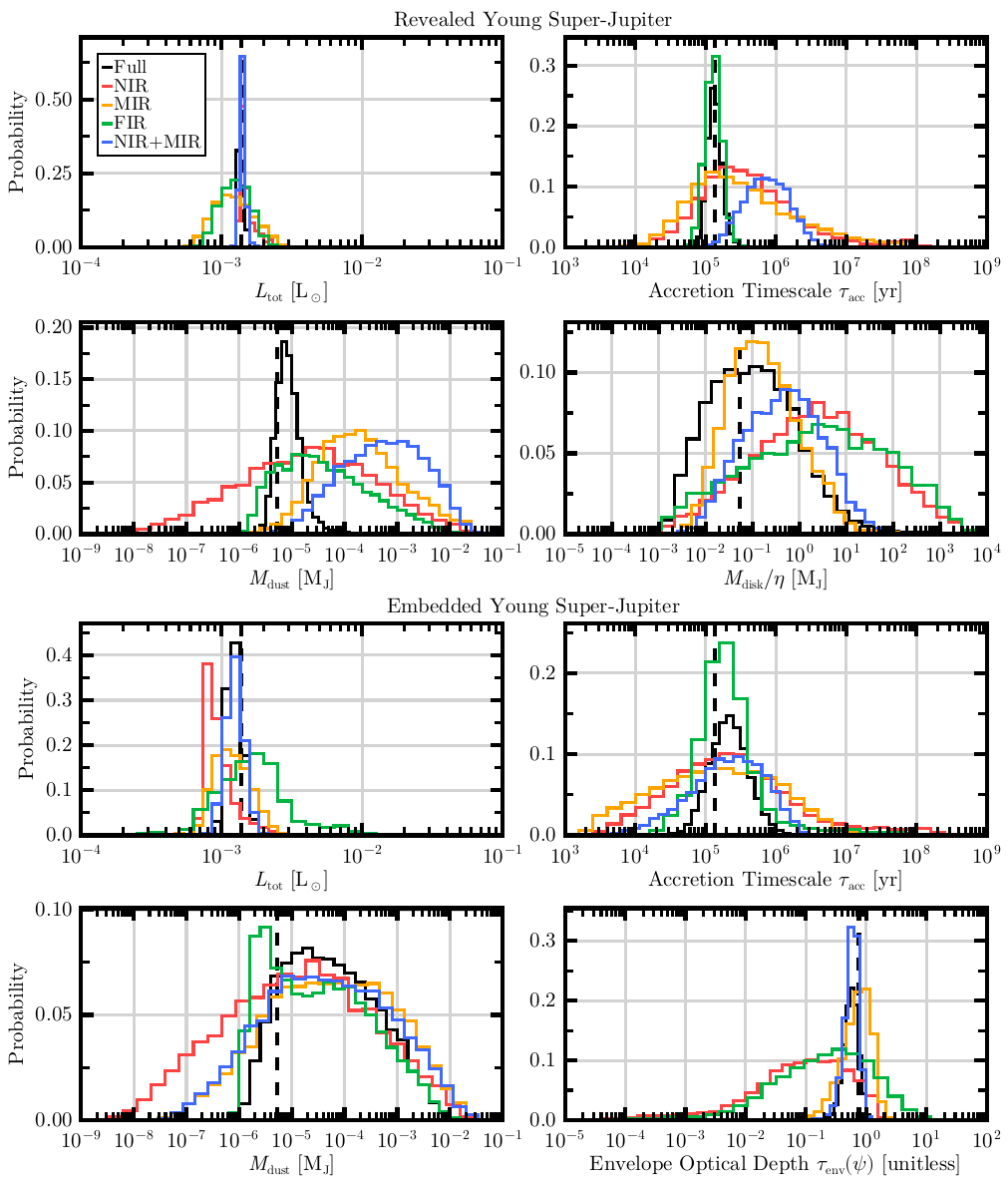}
    \vspace{-5pt}
    \caption{\textbf{Young Super-Jupiter Derived Parameters.} Posteriors of several important parameters for observations of a revealed and an embedded young super-Jupiter over a range of wavelengths. The total luminosity $L_{\rm tot}$ and the accretion timescale $\tau_{\rm acc}$ are linearly independent and jointly define $M_p$ and $\dot{M}$ and $M_{\rm dust}$ and $M_{\rm disk}/\eta$ jointly define $\alpha$ and $\eta$. For a revealed system, observations are capable of putting tighter constraints on $L_{\rm tot}$ and $M_{\rm dust}$ than on $\tau_{\rm acc}$ or $M_{\rm disk}/\eta$. For an embedded system, the dust mass is poorly constrained, although the envelope optical depth $\tau_{\rm env}$ is reasonably well-constrained. The colors show the fits for observations in different wavelength bands---the NIR (red), MIR (orange), FIR (green), combined NIR and MIR (blue), and the full SED (black). }
    \vspace{-15pt}
    \label{fig:YSJhists}
\end{figure*}

\subsection{Young Super-Jupiter}

The properties of a young super-Jupiter are considered next. Similarly to the young Jupiter, the constraints are stronger on derived parameters ($L_{\rm tot}$, $\tau_{\rm acc}$, etc) than on system parameters directly. For brevity, we will not show the full cornerplots and only present posteriors of the derived parameters (Fig.~\ref{fig:YSJhists}). Since this system is brighter, the full SED constrains the luminosity to $\Delta L_{\rm tot}\simeq\qty{0.16}{dex}$, although the constraints on the other parameters achieve the same order of magnitude as the young Jupiter model. The information contained in each wavelength band is the same for both the young super-Jupiter and the young Jupiter models.

The posterior of the embedded young super-Jupiter model shows similar qualitative behavior to that of the young Jupiter. The constraints on $L_{\rm tot}$ and $\tau_{\rm acc}$ have similar precision and accuracy to the constraints imposed by observations of a revealed system, and the constraints on $M_{\rm dust}$ are slightly weaker. The NIR and joint NIR/MIR observations provide the strongest constraints on $L_{\rm tot}$ and the FIR provides the best constraint on $\tau_{\rm acc}$. Multi-band observations are capable of constraining the envelope optical depth to within \qty{0.35}{dex}, with the MIR and joint NIR/MIR observations providing the strongest constraints (aside, of course, from the full SED). 

Other than these small differences, the parameter constraints for these systems are nearly identical. Although both the young Jupiter and young super-Jupiter have similar planet masses, this cannot fully explain the nearly identical parameter constraints. After all, the SEDs of these objects differ significantly (see Fig.~\ref{fig:exempmodelcomp}), although their SEDs have a similar shape. Instead, this agreement is likely a result of the optical depth of the circumplanetary disk. For both of these models, the disk is optically thick throughout, so that the information contained within a given observation is essentially equivalent. The mature super-Jupiter, with a much lower mass accretion rate, is optically thin at the outer edge of the disk and so exhibits different constraints.

As for the young Jupiter, the FIR contains the most information about the accretion timescale, since the location and temperature of the outer edge of the disk depends on the mass and mass accretion rate individually. The FIR does not provide a similar constraint for circumstellar disks, where the outer radius is not as strongly dependent on the stellar mass and the mass accretion rate. In the circumplanetary context, the outer edge of the disk is truncated by the centrifugal radius, while the outer edge of the circumstellar disk depends on the size of the primordial protostellar cloud. The absence of an FIR instrument thus makes it difficult to measure the mass and mass accretion rates of forming giant planets. 

\begin{figure*}
    \centering
    \vspace{-30pt}
    \includegraphics{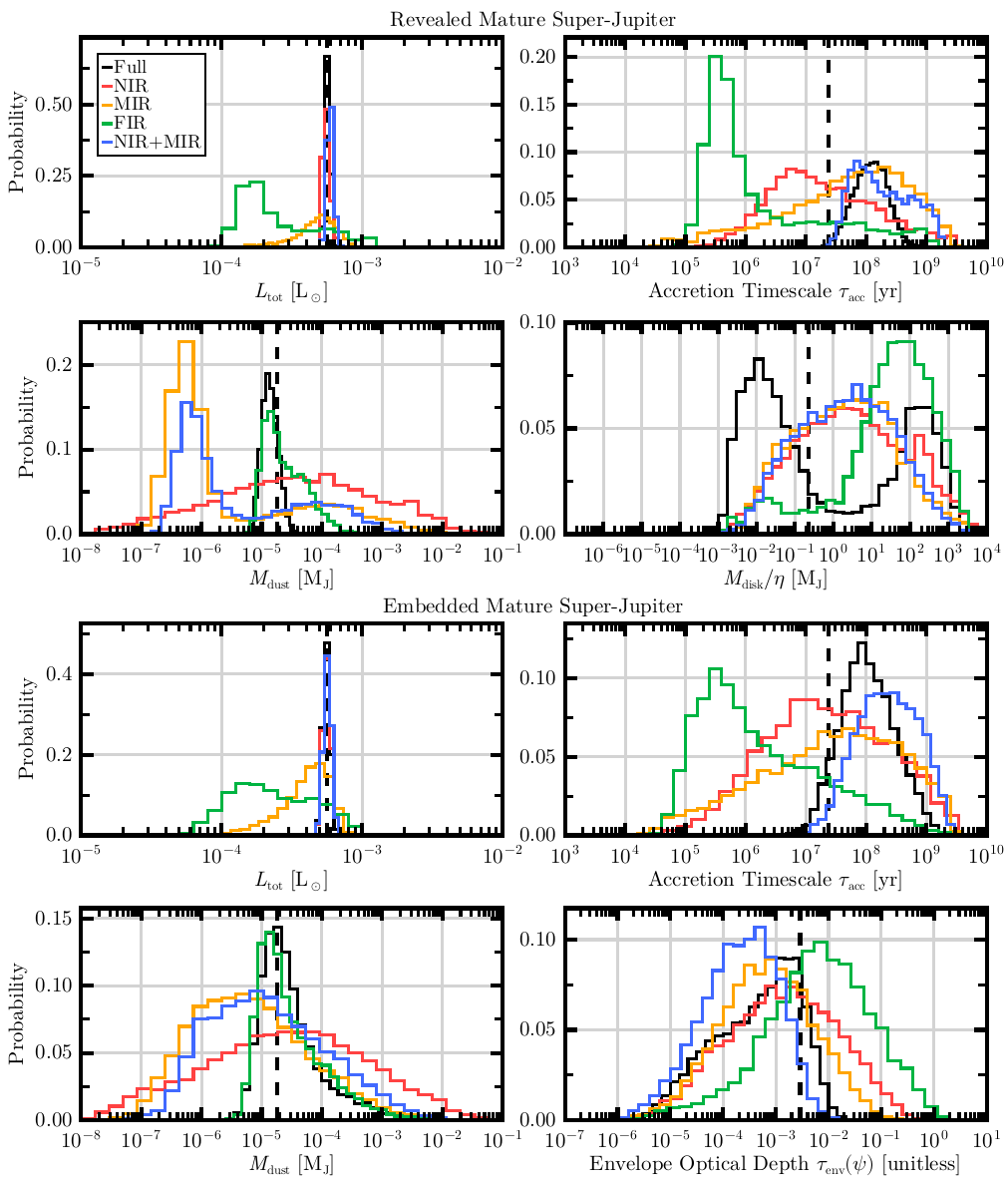}
    \vspace{-5pt}
    \caption{\textbf{Mature Super-Jupiter Derived Parameters.} Posteriors of several important parameters for observations of a revealed and an embedded mature super-Jupiter over a range of wavelengths. The total luminosity $L_{\rm tot}$ and the accretion timescale $\tau_{\rm acc}$ are linearly independent and jointly define $M_p$ and $\dot{M}$ and $M_{\rm dust}$ and $M_{\rm disk}/\eta$ jointly define $\alpha$ and $\eta$. For a revealed system, observations are capable of putting tighter constraints on $L_{\rm tot}$ and $M_{\rm dust}$ than on $\tau_{\rm acc}$ or $M_{\rm disk}/\eta$. For an embedded system, the dust mass is not as well constrained while observations put an upper limit on the envelope optical depth $\tau_{\rm env}$. The colors show the fits for observations in different wavelength bands---the NIR (red), MIR (orange), FIR (green), combined NIR and MIR (blue), and the full SED (black). }
    \vspace{-15pt}
    \label{fig:MSJhists}
\end{figure*}

\subsection{Mature Super-Jupiter}

We finally consider the parameter constraints for a mature super-Jupiter, with the derived parameter posteriors shown in Fig.~\ref{fig:MSJhists}. Once again, the constraints are somewhat tight on $L_{\rm tot}$ ($\Delta L_{\rm tot}\simeq\qty{0.05}{dex}$ for a revealed system and $\Delta L_{\rm tot}\simeq\qty{0.075}{dex}$ for an embedded system). Since the mass accretion rate is low, the planet emission in the NIR dominates the SED and the NIR constraint on the total system luminosity is slightly tighter for the mature super-Jupiter than for the other example models. 

While the constraints on $L_{\rm tot}$ are comparable to that of the young systems, $\tau_{\rm acc}$ is very poorly constrained ($\Delta\tau_{\rm acc}=\qty{0.8}{dex}$ and $\Delta\tau_{\rm acc}=\qty{1.35}{dex}$ for revealed and embedded systems). This behavior can be almost entirely attributed to a change in the information delivered by the FIR, which provides significantly weaker constraints on $\tau_{\rm acc}$. Unlike the other models, the mature super-Jupiter disk is optically thin in its outer regions, which are observed in the FIR. For an optically thin disk, the scale height of the outer disk has no real effect on the SEDs, so the constraints on $M_p$ and $\dot{M}$ individually discussed above are weak, and there are a range of total luminosities that can reproduce the total flux in the FIR. So long as the outer disk is optically thin, many values of $M_p$ and $\dot{M}$ match the data. 

Meanwhile, the FIR provides tighter constraints on the dust mass for the mature super-Jupiter in comparison to the other models. In fact, most of the dust-mass information contained in the entire SED is contained in the FIR, for both revealed and embedded planets. For an optically thin outer disk, the dust mass has a significant impact on the FIR luminosity (so long as the opacity is fixed, as it is here). For disks that are optically thick throughout (such as the young Jupiter and young super-Jupiter models), the full SED is necessary to estimate $M_p$ and $\dot{M}$ and to approximately constrain the total dust mass. For this system, only the FIR is necessary. The constraint on $M_{\rm dust}$ provided by the full SED of a revealed system is approximately \qty{0.4}{dex} for the mature super-Jupiter, noticeably smaller than the $\sim\qty{1}{dex}$ constraint for the younger planet models. 

Since the mass accretion rate of this system is small, the circumplanetary envelope has little impact on the system SEDs (cf. Fig.~\ref{fig:exempmodelcomp}). Even with full spectral coverage, observations of these systems cannot constrain the total envelope optical depth to less than $\Delta\tau_{\rm env}\simeq\qty{2.5}{dex}$. This is in sharp contrast to the constraints on $\tau_{\rm env}$ for an embedded young system, which are at most $\Delta\tau_{\rm env}\simeq\qty{0.5}{dex}$. Since the envelope is optically thin with $\tau_{\rm env}\simeq\num{3e-3}\ll1$, observations can only provide an upper limit on the optical depth. 

The tapering off of the posterior at low optical depths can be attributed to the fact that such low optical depths require a very large mass and a very small mass accretion rate to match the system luminosity. Such a system would have a geometrically thinner disk and an SED that is a poor match for the synthetic observations. 

Since the SED is effectively unchanged, the constraints on $L_{\rm tot}$, $\tau_{\rm acc}$, and $M_{\rm dust}$ are mostly equivalent to the constraints for a revealed system. Unlike the young systems, $M_{\rm dust}$ is equally well-constrained for a revealed or embedded mature super-Jupiter. The additional degeneracies introduced by the envelope absorption weaken the constraints somewhat, but only slightly. Although it will be difficult to constrain the total optical depth of such a thin envelope, multiband or FIR observations are still able to provide a \qty{1.5}{dex} constraint on the total dust disk mass.

These results reveal that there is a qualitative difference between the constraints that can be placed on a system with an optically thin circumplanetary disk versus a disk that is optically thick throughout. For completely optically thick disks (e.g., the young Jupiter and super-Jupiter models), the total luminosity can be constrained to within approximately \qty{0.05}{dex} while the accretion timescale can be constrained to \qty{0.35}{dex} if full wavelength coverage is available. On the other hand, the total system dust mass and $M_{\rm disk}/\eta$ are only loosely constrained. 

If the outer disk is optically thin, then the accretion timescale is only loosely constrained ($\Delta\tau_{\rm acc}\gtrsim\qty{0.8}{dex}$), but the FIR allows for measurements of the dust mass. 

The low mass accretion rates necessary for a system to have an optically thin outer disk generally mean that the envelope would also be optically thin. If these systems are embedded in the circumstellar disk, the dust mass is still constrained, although the measured envelope optical depth can span several orders of magnitude. 

\subsection{Observability}

While the previous sections show the constraints on planet properties that can be obtained from observations of an embedded protoplanet, these fits assume that the input SEDs capture only the energy emergent from the Hill sphere of the protoplanet. For a system embedded in a circumstellar disk, the radiative signatures are further attenuated by the intervening disk, which may render the protoplanet undetectable.

For a minimum-mass solar nebula, the column density to the midplane is \citep{Hayashi1981}
\begin{equation}
    N_{\rm col}=\qty{876}{g/cm^2}\left(\frac{a}{\unit{au}}\right)^{-3/2}\,.
\end{equation}
Further attenuating all of the example model SEDs by this column density and the assumed total opacity almost completely obscures the radiative signatures. Correcting for this additional absorption is complex. Even if such a planet can be detected, the column density between the observer and the protoplanet's Hill sphere depends on the size, distribution, and composition of the circumstellar disk dust, which is difficult to measure. In addition, if the optical depth of the circumstellar disk is large, then the absorption and reradiation of the protoplanet's radiation would completely obscure all information contained within the SED. Such observations are highly unlikely to be successful, and only objects in deep gaps are likely to be observable. 

\begin{figure}
    \centering
    \includegraphics{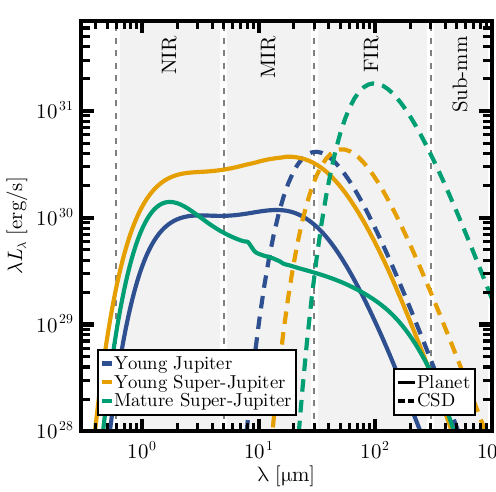}
    \caption{\textbf{Comparing Circumstellar Disk SEDs.} A comparison of the SEDs that emerge from the Hill sphere of the example protoplanet models (solid) versus the SEDs of the background circumstellar disk (dashed). The protoplanet SEDs are calculated by \RADp, while the circumstellar disk SEDs are calculated using Eqs. \eqref{eq:TCSD} and \eqref{eq:LCSD}. }
    \vspace{-10pt}
    \label{fig:CSDcomp}
\end{figure}

Detections must also be able to distinguish a forming planetary system from the background circumstellar disk. The SEDs of a $\pi R_H^2$ region of the background circumstellar disk, which the protoplanetary system essentially replaces, are compared to the circumplanetary system SEDs in Fig. \ref{fig:CSDcomp}. The temperature of the background disk takes the form 
\begin{equation}\label{eq:TCSD}
    T_{\rm CSD}=T_{X, \rm CSD}\left(\frac{a}{R_{X, \rm CSD}}\right)^{-1/2}\,.
\end{equation}
In this equation, $T_{X, \rm CSD}$ is the dust destruction temperature at \qty{1500}{K}, while $R_{X, \rm CSD}=\qty{0.04}{au}$ is the distance at which the equilibrium temperature for a \qty{1.3}{L_\odot} star is equal to \qty{1500}{K}. A circumstellar disk can have a wide range of possible temperature profiles, so this is merely an illustrative example. The luminosity of the circumstellar disk is then given by 
\begin{equation}\label{eq:LCSD}
    L_{\nu, \rm CSD}=4\pi\,\pi R_H^2\,B_\nu(T_{\rm CSD})\,,
\end{equation}
where $4\pi$ is a normalization convention and the factor of $\pi R_H^2$ accounts for the surface area of the disk that is replaced by the protoplanet system. 

Fig. \ref{fig:CSDcomp} shows that in general, the protoplanet is significantly hotter and bluer than the background disk and can be distinguished from the background. Of course, the contrast is even larger if the protoplanet is embedded in a gap, which is the most likely scenario fo detecting these objects.

\begin{figure}
    \centering
    \includegraphics{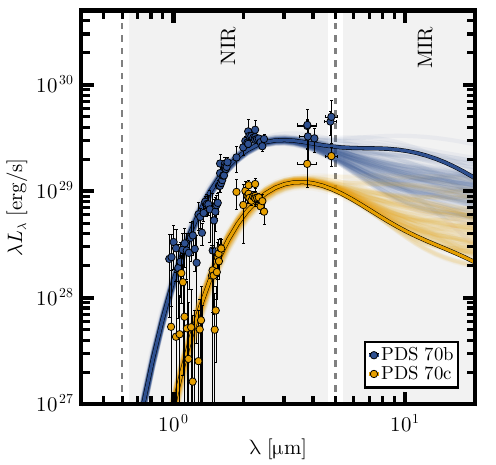}
    \vspace{-10pt}
    \caption{\textbf{PDS 70 SEDs.} The observed NIR SEDs and photometry of PDS 70 planets b and c (points) along with fit SEDs. The solid lines are the maximum-likelihood SEDs while the transparent lines are sampled from the posterior distribution.}
    \vspace{-10pt}
    \label{fig:PDS70}
\end{figure}

\section{Applications to Observed Systems}\label{sec:realplanets}

To demonstrate the usefulness of this methodology, these models are applied to observations from the literature of several detected circumplanetary disks. In particular, we study the SEDs of PDS 70 b and c, the first-detected and most well-studied circumplanetary disk system. The SED of GQ Lup b, a widely-separated substellar object, is also studied. While SEDs of other systems exist --- e.g., WISPIT 2b \citep{Close2025a} --- these objects generally only have a few photometric measurements, making robust parameter constraints infeasible.  {The constraints derived in this section are not absolute but rather are specific to the assumptions of this model.} 

\subsection{PDS 70 Protoplanets}

PDS 70 itself is a \qty{0.76}{M_\odot} K dwarf and T Tauri system located \qty{112}{pc} away \citep{Keppler2018}. It hosts two confirmed planets, PDS 70 b and PDS 70 c \citep{Isella2019,Benisty2021}. PDS 70 b has an orbital distance of $a\simeq\qty{20.8}{au}$ while PDS 70 c orbits at $a\simeq\qty{34.3}{au}$ \citep{Wang2021}. {The orbits of the planets --- and by assumption the angular momentum vectors of the circumplanetary disks --- are} inclined by approximately $\psi=\qty{51.7}{\degree}$, the same as the circumstellar disk \citep{Keppler2019, Wang2021}. The planet radii are both set to $R_p\simeq\qty{2}{R_J}$, as inferred from model retrievals \citep{Wang2021}. This value is also expected to be the radius of the young Jupiter \citep{Batygin2025, Knierim2026}. 

Observations of these objects have been made using a wide variety of instruments, including VLT/SPHERE \citep{Muller2018, Mesa2019}, VLTI/GRAVITY \citep{Wang2021}, Keck/NIRC2 \citep{Wang2020}, Gemini/NICI \citep{Muller2018}, VLT/NaCo \citep{Stolker2020}, JWST/NIRCam \citep{Christiaens2024a}, and ALMA \citep{Isella2019, Benisty2021, Dominguez-Jamett2025}. Except for ALMA, these data are exclusively in the band $\lambda=\qtyrange{0.6}{5}{\micro\meter}$ and fall in the NIR. The ALMA data are highly sensitive to the opacity of the disk dust and will be ignored. While there have been MIR observations of the PDS 70 system \citep{Perotti2023, Jang2024}, these observations do not spatially resolve the protoplanets and cannot be used here.

\renewcommand{\arraystretch}{1.25}
\begin{table}[]
    \centering
    \caption{\textbf{Calculated Parameters.} The parameter values calculated for the PDS 70 protoplanets and GQ Lup b using the SAM. The GQ Lup b values are reported for the model without an envelope. Including an envelope returns nearly identical results. This table reports the median and \qty{68}{\percent} confidence interval values for each parameter. }
    \begin{tabular}{l|l|l|l}
        Parameter & PDS 70 b & PDS 70 c & GQ Lup b \\\hline
        $M_p$ [\unit{M_J}] & $4.37^{+13.3}_{-3.06}$ & $4.59^{+11.2}_{-3.5}$ & $28.97^{+30.19}_{-15.75}$\\
        $\dot{M}$ [\unit{M_J/Myr}] & $0.21^{+0.45}_{-0.15}$ & $0.08^{+0.25}_{-0.06}$ & $1.21^{+1.38}_{-0.62}$ \\
        $B_{p,0}$ [\unit{G}] & $580^{+270}_{-300}$ & $660^{+220}_{-270}$ & $280^{+320}_{-140}$\\
        $f_{\rm Si}$ & $0.55^{+0.34}_{-0.34}$ & $0.55^{+0.34}_{-0.34}$ & $0.66^{+0.29}_{-0.36}$\\
        $\alpha$ [\num{e-3}] & $12^{+170}_{-11}$ & $14^{+160}_{-13}$ & $121^{+576}_{-117}$ \\
        $\eta$ [\num{e-3}] & $2.9^{+20}_{-2.5}$ & $2.4^{+19}_{-2.1}$ & $2.8^{+18}_{-2.6}$ \\
        $\psi$ [\unit{\degree}] & - & - & $60.1^{+6.1}_{-14.7}$\\\hline
        $L_{\rm tot}$ [\qty{e-4}{L_\odot}] & $1.57^{+0.09}_{-0.06}$ & $0.65^{+0.04}_{-0.03}$ & $50.8^{+5.0}_{-4.5}$ \\
        $\tau_{\rm acc}$ [\unit{Myr}] & $21^{+310}_{-19}$ & $58^{+620}_{-55}$ & $24^{+76}_{-19}$ \\
    \end{tabular}
    \label{tab:PDS70pars}
\end{table}

The MCMC is used to fit the SAM to the data in the literature. As before, we use \num{30} walkers, a \num{1000} step burn-in, and \num{100000} sample steps. The values of $M_p$, $\dot{M}$, $B_{p,0}$, $f_{\rm Si}$, $\eta$, and $\alpha$ are free parameters while $R_p$, $\psi$, $a$, and $M_\star$ are fixed. Since the PDS 70 objects are known to reside in a deep gap, the extinction is assumed to be zero and the circumplanetary envelope is turned off. Fig.~\ref{fig:PDS70} shows the data points along with the maximum-likelihood SEDs and SEDs sampled from the posterior. 

For the PDS 70 protoplanets, the maximum-likelihood SEDs accurately match the observations at $\lambda\leq\qty{4}{\micro\meter}$, although the SAM under-predicts the luminosity of the VLT/NaCo photometry at \qty{4.5}{\micro\meter}. The brightness of these photometric points may be the result of undetected gas emission features, in particular a CO line at \qty{4.25}{\micro\meter} and/or a broad CO$_2$ feature at \qty{4.5}{\micro\meter} \citep{McClure2025}. Without MIR data, the SED posterior becomes highly unconstrained at those wavelengths. Additional observations in the MIR or FIR would be instrumental in clarifying the properties of these objects. 

The posterior distribution from these MCMCs has tight constraints on $L_{\rm tot}$ (\qty{0.075}{dex} for both PDS 70 b and c), but significantly looser constraints on all other parameters. In particular, the accretion timescale $\tau_{\rm acc}$ is only constrained to $\sim$\qty{2.5}{dex}, while the other parameters are unconstrained in the parameter space. The median parameter values for PDS 70 b are $M_p=\qty[parse-numbers  = false ]{4.37^{+13.3}_{-3.06}}{M_J}$, $\dot{M}=\qty[parse-numbers  = false ]{0.21^{+0.45}_{-0.15}}{M_J/Myr}$ and $M_p=\qty[parse-numbers  = false ]{4.59^{+11.2}_{-3.5}}{M_J}$, $\dot{M}=\qty[parse-numbers  = false ]{0.08^{+0.25}_{-0.06}}{M_J/Myr}$ for PDS 70 c, where the quoted errors are the 16th and 84th percentiles. The predicted values (with errors) of all parameters are given in Table \ref{tab:PDS70pars}. Except for $M_p$ and $\dot{M}$, the system parameters are effectively unconstrained.

The values of $M_p$ and $\dot{M}$ are relatively close to those predicted by other authors. First, the PDS 70 c value of $M_p\dot{M}\simeq\qty{0.4}{M_J^2/Myr}$ is consistent with the lower limit derived from ALMA measurements of dust thermal emission \citep{Shibaike2024}. In particular, \qty{2}{\sigma} dynamical upper mass limits are $M_p\leq\qty{4.9}{M_J}$ for PDS 70 b and $M_p\leq\qty{13.6}{M_J}$ for PDS 70 c \citep{Trevascus2025}, which are somewhat consistent with these values. The mass of PDS 70 b calculated by this work is at the edge of this \qty{2}{\sigma} upper limit, but is consistent with the reported value. These values are also consistent with other mass measurements, which predict that the masses of these planets are approximately \qty{10}{M_J} \citep{Muller2018, Aoyama2019}.

Measurements of the mass accretion rate of these planets based on their H$\alpha$ emission span a range of values. For PDS 70 b and PDS 70 c, the accretion rates are on the order of $\dot{M}\simeq\qty{e-2}{M_J/Myr}$ \citep{Wagner2018, Haffert2019, Aoyama2019, Zhou2021}, $\dot{M}\leq\qty{e-3}{M_J/Myr}$ \citep{Thanathibodee2019}, or $\dot{M}\simeq\qty{e-4}{M_J/Myr}$ \citep{Close2025}. While our estimated accretion rate for PDS 70 c is consistent with the upper range of these values, our median accretion rate for PDS 70 b is approximately an order of magnitude too large. The significant uncertainties in these values are the result of different model assumptions and uncertainty in the ({small}) H$\alpha$ luminosity. 

The H$\alpha$-derived mass accretion rates rely heavily on assumptions of the dust extinction between the circumplanetary system and the observer due to either the circumplanetary envelope or the remaining material in the circumstellar disk gap. The extinction in these gaps may be larger than expected \citep{Cugno2025b}, requiring higher accretion rates to explain the flux. It has also been shown that the line-emitting accretion rate is generally a fraction of the true accretion rate, implying that the true accretion rates are much larger \citep{Marleau2022, Marleau2023, Marleau2025}. Although we also assume zero dust extinction between the Hill sphere and the observer, this assumption is more appropriate in the NIR than for H$\alpha$ band, since the opacity is lower in the infrared. If an extinction of $A_{H\alpha}\simeq\qty{2}{mag}$ is accounted for, the mass accretion rate of PDS 70 b is $\dot{M}\simeq\qty{0.5}{M_J/Myr}$ \citep{Hashimoto2020}, consistent with our measurements. {Assuming a typical ISM extinction law, an extinction of \qty{2}{mag} at the H$\alpha$ wavelength of \qty{0.66}{\micro\meter} corresponds to an extinction of \qty{0.33}{mag} at \qty{2}{\micro\meter}, maintaining $\sim$\qty{75}{\percent} of the flux (see \citealt{Cardelli1989}). This correction is minor relative to the local point scatter, and is even less significant at longer wavelengths.} These results imply that there may be significant extinction to PDS 70 b, albeit less to PDS 70 c.  

\begin{figure}
    \centering
    \includegraphics{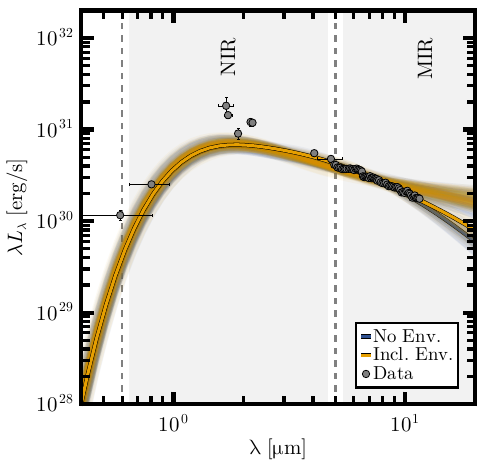}
    \vspace{-10pt}
    \caption{\textbf{GQ Lup b SEDs.} The observed NIR and MIR SEDs and photometry of GQ Lup b (points) along with fit SEDs. The solid lines are the maximum-likelihood SEDs while the transparent lines are sampled from the posterior distribution. The MCMC is run both including and excluding the circumplanetary envelope, and both SED posteriors are shown. }
    \vspace{-10pt}
    \label{fig:GQLupb}
\end{figure}

\subsection{GQ Lup Secondary}

Unlike the PDS 70 protoplanets, which are embedded in their natal disk, GQ Lup b falls outside the regime that the SAM is designed for. The orbital plane of GQ Lup b is misaligned from the circumstellar disk, so that the companion only passes through the disk twice over its orbit \citep{Wu2017, Venkatesan2025}. The accretion onto GQ Lup b from the circumstellar disk is therefore sporadic, if it occurs at all. Given this fact, the orbital parameters \citep{Meyer2025}, the compact circumstellar disk \citep{MacGregor2017}, and the chemical composition of the GQ Lup system \citep{GonzalezPicos2025}, it is likely that GQ Lup b formed through turbulent fragmentation of the protostellar cloud. The detected disk therefore may be the remnant of the protostellar cloud, rather than accreted material from the circumstellar disk. 

This qualitative difference means that the outer edge of the disk is not necessarily set by the size of the Hill radius, a crucial assumption in the SAM. However, this outer edge is mostly visible in the FIR and sub-mm, where observations either do not exist or are subject to external uncertainties. Although the SAM is not designed for such objects, this model still accounts for the geometric thickness and relative luminosity of these disks. Applying the SAM to objects like GQ Lup b can still provide insight into the properties of the system, although care must be taken when interpreting these results. 

GQ Lup A is a \qty[parse-numbers  = false ]{1.03\pm0.05}{M_\odot} T Tauri star in the Lupus I star forming region \citep{MacGregor2017}. There may be a binary M-dwarf companion at \qty{2400}{au} \citep{Alcala2020}. The substellar companion GQ Lup b \citep{Neuhaeuser2005} orbits at \qty{97.7}{au} with an orbital inclination of \qty{48}{\degree} \citep{Venkatesan2025}. 

Observations of GQ Lup b range from the optical to the MIR. SEDs and photometry have been obtained from HST/WFPC2, HST/NICMOS, and Subaru/CIAO \citep{Marois2006}, VLT/NaCo, VLT/MUSE and VLT/SINFONI \citep{Stolker2021}, JWST/MIRI \citep{Cugno2024a}, and MagAO and ALMA \citep{Wu2017}. Some of these SEDs (VLT/MUSE and VLT/SINFONI) are not easily accessible, and the ALMA data are fraught due to the opacity sensitivity. The rest of the data are used to fit the SAM to these observations. 

The MCMC parameters are the same as for the PDS 70 objects. The significant mutual inclination between the circumstellar disk and the orbital plane of GQ Lup b means that the envelope is unlikely to be present, although the envelope infall time is comparable to the orbital time. To test this, we have run MCMCs both including and excluding the circumplanetary envelope. The viewing angle $\psi$ of the GQ Lup b disk is also set to a free parameter, since the orbital geometry means that the circumplanetary disk may not be aligned with the circumstellar disk. The data are shown with the SED posterior in Fig. \ref{fig:GQLupb}

The SAM accurately reproduces the optical photometry and the MIR SED, but underpredicts the NIR luminosity from $\lambda=\qtyrange{1}{3}{\micro\meter}$. VLT/MUSE and VLT/SINFONI data in this wavelength band show significant spectral features \citep{Stolker2021}, which are likely responsible for this residual. This error could also be the result of the internal luminosity of the planet, which is assumed to be negligible in the SAM. These data tightly constrain $L_{\rm tot}$ to \qty{0.15}{dex} but only place \qty{2}{dex} constraints on $\tau_{\rm acc}$. 

The median values and \qty{68}{\percent} confidence intervals of the system parameters are given in Table \ref{tab:PDS70pars}. The silicate fraction, dust-to-gas ratio, magnetic field strength, and viscosity parameter are effectively unconstrained. However, the luminosity is constrained to within \qty{20}{\percent}, while the mass and mass accretion rate are constrained to within an order of magnitude. Similar to the derived parameters for the PDS 70 objects, our fit to GQ Lup b is broadly consistent with prior mass measurements (\qtyrange{10}{40}{M_J}, \citealt{Stolker2021} and the extinction-corrected H$\alpha$ measurements of $\dot{M}\simeq\qty{0.3}{M_J/Myr}$ \citep{Zhou2014, Stolker2020, Demars2023}.
Since the SAM does not account for the internal luminosity of the central source, either the mass or mass accretion rate may be inflated. Given the mass of this object and its young age \citep{Xuan2024}, the internal luminosity is likely significant. 

This fit also predicts a somewhat narrow range of likely viewing angles, between $\psi=\qtyrange{45}{65}{\degree}$. {While this inclination value is consistent with the orbital inclination of $48^{+3.7}_{-4.9}$ \unit{degree} \citep{Venkatesan2025}, the on-sky angle of the circumplanetary disk angular momentum vector is unconstrained, allowing for any mutual inclination between the circumplanetary disk and orbital angular momentum vectors.} This value is inconsistent with the highly inclined disk found by fits using thin-disk models \citep{Cugno2024a}. This discrepancy likely arises from the true geometric thickness of this disk, which the SAM accounts for. 

Finally, the relative NIR/MIR slope of this SED is inconsistent with a significant circumplanetary envelope. The close agreement between the SED posteriors both including and excluding the circumplanetary envelope (see Fig. \ref{fig:GQLupb}) and the agreement between the derived parameters also shows that the SED is inconsistent with significant attenuation. When the envelope is included, the envelope optical depth has an upper limit of $\tau_{\rm env}\lesssim\num{0.015}$ on the envelope optical depth, confirming this result.

\section{Conclusion}\label{sec:conc}

\subsection{Summary}

In this work, we have constructed a semianalytic model for the structure and radiative signatures of geometrically thick circumplanetary disks (Sec. \ref{sec:SAM}). This model provides a significant improvement over previous thin-disk models in matching numerical simulations of the structure and radiative signatures of these systems, while retaining similar computational efficiency. This model can be used with an MCMC to fit spectral observations of forming planets and their circumplanetary disks, and thereby constrain the properties of these systems ({subject to the assumptions inherent in the model}). 

Using this model, we have shown that multiband observations of forming giant planets can robustly measure the total system luminosity to better than \qty{0.1}{dex}. However, separating the planet mass from the mass accretion rate requires observations in the FIR and circumplanetary disks that are optically thick throughout  {(see also \citealt{Zhu2015})} . For optically thin disks, FIR observations only constrain the total dust mass of the disk. 

For a young protoplanet with a high mass accretion rate, the circumplanetary disk is entirely optically thick in the infrared. If this planet is accreting via a streamer, NIR observations will provide strict constraints on the planet luminosity, but not $M_p$ and $\dot{M}$ separately. Absent other spectral features, only observations in the FIR can break this degeneracy via measuring the accretion timescale. For circumplanetary disks, the location of the disk edge depends on the Hill radius and is sensitive to the mass and mass accretion rate separately, enabling this constraint. The fact that the FIR constrains these parameters is a convenient property of circumplanetary disks. Full-spectrum observations (NIR, MIR, and FIR) are also able to constrain the total dust mass of the system, but only to within an order of magnitude. 

If a young protoplanet is embedded within an envelope (and hence within the circumstellar disk), the observational constraints on the total luminosity and the accretion timescale are weaker. While observations of these objects cannot constrain the dust mass, joint NIR and MIR observations can measure the optical depth of the envelope. This result indicates that the slope of the NIR/MIR SED is able to determine if a significant envelope is present. The planet (which peaks in the NIR) is always brighter than the disk (which emits in the MIR). If the MIR is instead brighter than the NIR, then the planet must be attenuated by the envelope, which would re-emit in the MIR.

For systems with low mass accretion rates (i.e., mature systems) the luminosity and the dust mass can be constrained reasonably well by the NIR and the FIR respectively. However, the mass and mass accretion rate are difficult to constrain separately using continuum emission alone, even with FIR observations. Even if the envelope is included, the envelope optical depth is sufficiently low for the SEDs to be essentially unmodified. As a consequence, the envelope optical depth is difficult to constrain, with observations only providing an upper limit. On the other hand, since the envelope is so diffuse, observations including an envelope do not lose any significant precision. 

To demonstrate this framework, these models are applied to literature observations of circumplanetary disks. The canonical case of the PDS 70 protoplanets illustrates the luminosity constraint imposed by NIR data. The current set of observations constrains the luminosity to within fractions of a \unit{dex}, while the accretion timescale is only loosely constrained. The masses and mass accretion rates measured for these systems are therefore highly uncertain and span approximately an order of magnitude. 

Despite the large uncertainties, the median values calculated by this model ($M_p=\qty{4.37}{M_J}$ and $\dot{M}=\qty{0.21}{M_J/Myr}$ for PDS 70 b and $M_p=\qty{4.59}{M_J}$ and $\dot{M}=\qty{0.08}{M_J/Myr}$ for PDS 70 c) are broadly consistent with prior measurements from both dynamical mass measurements \citep{Muller2018, Aoyama2019, Wang2021, Trevascus2025} and H$\alpha$ accretion measures \citep{Wagner2018, Haffert2019, Aoyama2019, Thanathibodee2019, Zhou2021, Close2025}, if there are \qty{2}{mag} of extinction to PDS 70 b \citep{Hashimoto2020}. These results therefore suggest that there is some extinction to PDS 70 b, with PDS 70 c likely in a deep gap with little extinction.

While GQ Lup b falls outside the primary regime of the SAM, fits to this system demonstrate the information that can be extracted from observations of widely-separated objects. While there are once again significant uncertainties, this object is likely more massive than the PDS 70 planets and accreting at a slightly higher rate. The inferred mass and mass accretion rate of GQ Lup b ($M_p\simeq\qty{29.0}{M_J}$ and $\dot{M}\simeq\qty{1.2}{M_J/Myr}$) are also consistent with previously measured values, which implies that there is little extinction to this object.
We also predict an inclination of $\psi\simeq\qty{60}{\degree}$ between the observer and the circumplanetary disk normal, which is relatively well-constrained. 
Given the mass and orbital geometry of GQ Lup b and the properties of the circumstellar disk, GQ Lup b is likely to have only modest mass accretion from the background disk and is thus unlikely to have significant attenuation from a circumplanetary envelope. 
Fits including such an envelope predict an envelope optical depth of only $\tau\leq\num{0.015}$.
%, confirming that the local attenuation is small. 

% All of the mass accretion rates calculated by these models are negligible for the formation process. Equivalently, the accretion timescale for all three objects is approximately an order of magnitude larger than the \qtyrange{3}{5}{Myr} ages of PDS 70 or GQ Lup. While the H$\alpha$ accretion rates may be systematically underestimated due to unaccounted-for extinction, the SAM mass accretion rate measurements primarily rely on the NIR and MIR where extinction is weaker. For GQ Lup b, the NIR/MIR slope indicates that the mass accretion rate is indeed dynamically small. Implicitly, the mass accretion rate must have been higher in the past. Since an object with a higher mass accretion rate will have a higher luminosity and easier to detect, it is likely that detection can only occur once the mass accretion rate drops off as the planet clears a gap. 

\subsection{Discussion}

This work has characterized the effectiveness of future observations in constraining the properties of accreting protoplanets. We have found that even low-resolution SEDs or photometry are able to provide significant constraints on the properties of forming systems (see also \citealt{Zhu2015, Krieger2022}). Broad wavelength coverage is necessary to optimally constrain the properties of planetary systems. If full-spectrum observations are available, the mass and mass accretion rate can typically be constrained to within \qty{0.5}{dex}. At this precision, the relationship between mass and mass accretion rate can be roughly clarified by observations. Constraints at this precision may also help resolve the timescale of giant planet formation to within a factor of 3. 

Importantly, the FIR continuum contains information about the mass and mass accretion rate that are not present in any other wavelength band. New FIR imaging, spectral, and photometric capabilities would not only constrain the dust mass (as can be done for circumstellar disks and optically-thin circumplanetary disks), but also break the mass/mass accretion rate degeneracy for revealed young planets. This work therefore motivates the construction of future high-angular-resolution FIR capabilities. While the proposed PRIMA spacecraft \citep{Glenn2025a} will be able to provide these observations for widely-separated objects, it will have insufficient angular resolution to resolve protoplanets from the background disk. If the protoplanet is embedded in a sufficiently wide gap, then this detection may be possible with future instruments.

This work also motivates high-angular-resolution MIR observations of the PDS 70 system. As the only confirmed embedded protoplanets with circumplanetary disks, these objects provide crucial insights into the formation process of giant planets.
% \footnote{While other planetary-mass objects with disks have been detected, e.g. GQ Lup b \citep{Cugno2024a, }, these objects are generally on very wide orbits ($\sim\qty{100}{au}$) and likely formed via protostellar core fragmentation rather than core accretion in a disk \citep{Meyer2025}. This model is not applicable to these objects.} 
As current observations of these objects are limited to the NIR, only weak constraints on the parameters of these planets can be made. 
% Detections in the MIR will allow for an order-of-magnitude improvement on measurements of the critical properties of these planets. 
% 
Our fits to PDS 70 b and PDS 70 c predict that the typical MIR specific flux from $\lambda=\qtyrange{5}{14}{\micro\meter}$ is \qtyrange{8.1e-17}{1.3e-16}{W/\meter^2/\micro\meter} for PDS 70 b and \qtyrange{3.8e-17}{5.87e-17}{W/\meter^2/\micro\meter} for PDS 70 c. Unfortunately, these observations will require angular resolution that cannot be achieved by current instruments. The in-development MIRAC-5 instrument on MMT \citep{Bowens2022} and the forthcoming ELT/METIS spectrograph \citep{Brandl2021} may be able to detect these objects and thereby constrain the properties of these objects. 

As discussed in Sec. \ref{sec:realplanets}, the best-fit models to the PDS 70 protoplanet SEDs under-predict the luminosity at \qty{4.5}{\micro\meter}. Since this photometric point is at the edge of the domain, it is difficult to know if this disagreement is due to a fundamental model inaccuracy or spectral features (such as CO$_2$ lines) that are absent in the SAM. If the luminosity continues to rise in the MIR, then the SAM is either inaccurate even for continuum emission or these protoplanets are embedded in a surprisingly large envelope and have a larger-than-expected accretion rate. If the MIR luminosity instead drops off to levels predicted by our model, then this \qty{4.5}{\micro\meter} spike may be the result of gas spectral features, which would provide further information about the circumplanetary system. 

These calculations have assumed that the circumplanetary environment of the PDS 70 system is in a steady state. However, recent work has detected variable accretion in the PDS 70 system, with flux variability of a factor of 2 within two years \citep{Zhou2025}. While this variability is relatively small, it is possible that variable accretion accounts for the relatively high \qty{4.5}{\micro\meter} observation. 

Although our fits to the GQ Lup b SED are relatively accurate, they underestimate the luminosity at $\lambda\simeq\qty{2}{\micro\meter}$. These photometric points capture spectral features from the central planet and therefore are difficult for these models to fit. The SAM also assumes that the internal luminosity of the companion is small compared to the accretion luminosity, which is not necessarily true for this young and high-mass object. 

Our results may also explain the non-detections of circumplanetary material in ALMA surveys (e.g., \citealt{Andrews2021}). Out of the three fiducial models explored here, the young Jupiter and super-Jupiter models have total dust masses of $M_{\rm dust}\simeq\qty{e-4}{M_\oplus}$, generally less than the $\sim\qty{e-3}{M_\oplus}$ lower limits established by these surveys. Only the mature super-Jupiter, which is comparable to the ALMA-detected PDS 70 planets, has a dust mass above this cutoff. 

% SAM is also difficult to apply in the context of GQ Lup b. This model is designed to model objects that are embedded in their natal circumstellar disks, which means that the mass accretion sets the outer edge of the circumplanetary disk. Given the high mutual inclination between the orbit of GQ Lup b and the circumstellar disk, the outer edge of the disk is not necessarily set by the accretion flow. Since this object is unlikely to be accreting from the circumstellar disk, the detected disk is instead likely to be a remnant of a cloud fragmentation process. This also explains the relatively small size of the dusty disk --- the dust is likely drifting inwards radially and is not being replenished by the circumstellar disk. 

Critically, this work {provides further evidence }that the mass accretion rates of these objects are small in the sense that the accretion timescales (\qtyrange{20}{60}{Myr}) are much longer than the system ages. 
% While these observations do not break the degeneracy between the mass and the mass accretion rate and there is significant uncertainty, there are three reasons to suspect that the mass accretion rate is truly negligible. First, even the maximum mass accretion rate allowed by these observations is small. Second, if independent measurements of the mass of these objects (either atmospheric retrievals or dynamical limits) are used to break the degeneracy, the mass accretion rates are more tightly constrained to their current order of magnitude. Finally, the SAM does not account for the internal luminosity of the planet and assumes that all emission is due to the accretion luminosity. Since the total luminosity is well-constrained, any internal luminosity must come at the expense of the accretion luminosity. This assumption therefore means that the estimated mass accretion rate is a maximum. These results therefore confirm that the mass accretion rates of these protoplanets are too small to form the planet within the age of the system. 
This result was already suggested by H$\alpha$ observations, but low H$\alpha$ emission may be attributed to unaccounted-for optical extinction rather than genuinely small mass accretion rates. The measurements of $\dot{M}$ in this work rely on the NIR and MIR, where dust extinction is less significant {by factors of a few}. The fact that the measured $\dot{M}$ values here remain {small provides} evidence that the mass accretion rates are truly small and are not significantly underestimated. {Although the values of $\tau_{\rm acc}$ from this model are consistent with the age of the system, reaching agreement requires relatively low planet masses and high mass accretion rates and is therefore unlikely. This finding provides further evidence that} mass accretion rates of these planets higher in the past in order to produce the observed planets.

Since a higher accretion rate would mean a higher luminosity, a rapidly-accreting protoplanet should be more detectable. It is therefore likely that rapid accretion reduces detectability through some other mechanism. It may be the case that high accretion rates only occur when a planet is deeply embedded in the natal circumstellar disk, which strongly attenuates the luminosity and prevents detection. In this scenario, when the planet carves a gap and becomes observable, the accretion rate also drops off. The distribution of giant planet masses would thus be set primarily (or at least in part) by the gap-carving mass. It could also be the case that the accretion is episodic and these objects are observed at quiescent times, while active accretion is rare. Investigation of these scenarios is necessary to determine how giant planets form and evolve. 

These results can also be used to roughly constrain the initial entropy of these protoplanets (see \citealt{Spiegel2012, Helled2014}). Eqs. (9) and (13) in \cite{Marleau2014} provide the internal luminosity of a given protoplanet as a function of the planet's mass, age, and initial entropy. {Given the wavelength range of the observations, the measured luminosity is effectively the total luminosity.} We tentatively assume that {this luminosity} is dominated by the accretion flow, so that the internal luminosity is at most \qty{10}{\percent} of the total system luminosity. From Table \ref{tab:PDS70pars}, PDS 70 b has $L_{\rm int}\leq\qty{1.6e-5}{L_\odot}$, PDS 70 c has $L_{\rm int}\leq\qty{6.5e-6}{L_\odot}$, and GQ Lup b has $L_{\rm int}\leq\qty{5.1e-4}{L_\odot}$. Using the median SAM masses and approximate measured ages, we find that PDS 70 b has $S\leq\qty{9.6}{k_B/baryon}$, PDS 70 c has $S\leq\qty{9.2}{k_B/baryon}$, and GQ Lup b has $S\leq\qty{10.0}{k_B/baryon}$. This constraint depends on both the planet mass and the fraction of the total luminosity that can be ascribed to the internal luminosity of the planet. In the limiting case where $L_{\rm tot}\simeq L_{\rm int}$, all three of these objects can be attributed to hot-start formation. However, for the preferred models derived here, these initial entropy limits are inconsistent with the hottest starts but match warm- or cold-start formation (\citealt{Spiegel2012}, see also \citealt{Knierim2026}, which shows that a warm start is likely for Jupiter).  

These constraints are loose and approximate, as this calculation is a proof-of-concept. In the future, it would be useful to include $L_{\rm int}$ as a free parameter in the SAM{, rather than assuming that the internal luminosity is small}. This would allow MCMCs to derive upper limits on the internal luminosity and to separate the internal and accretion luminosities. These upper limits (and planet masses) could then be combined with planet evolution models and age estimates to derive estimates and constraints on the initial planetary entropy.

There are several assumptions inherent in these models that must be discussed. First, we assume a power-law opacity with a fixed normalization for both the silicate and carbon components. While the power law is reasonably appropriate for $\lambda\leq\qty{300}{\micro\meter}$ (except for around the \qty{10}{\micro\meter} feature), the normalization used here is dependent on the dust size distribution. The actual size of the dust in a given disk --- and the corresponding opacity normalizations --- are highly uncertain. Modeling ALMA data is therefore fraught with difficulty, since the emission in the sub-millimeter is almost certainly optically thin. 

This model also only considers the continuum dust opacity and does not include any additional spectral features. In the \RADp model, the dust opacity does include the \qty{10}{\micro\meter} spectral feature and the long-wavelength dropoff past \qty{100}{\micro\meter}, but this is not captured by the SAM. Neither \RADp nor the SAM model any gas features, which may prove to be important diagnostics of planet formation processes. Modeling the gas chemistry in circumplanetary disks and their radiative signatures will become increasingly important as more detections of these objects are made using gas features (e.g., \citealt{Bae2022, Izquierdo2026}) and will be investigated in future work. 

% If the properties of the circumplanetary system can be tightly constrained from infrared observations, then extending the SED to the sub-millimeter would measure the opacity. Unfortunately, the results of Sec. \ref{sec:synthobv} show that constraining the relevant parameters ($\alpha$ and especially $\eta$) is difficult unless either the outer disk is optically thin (in which case the FIR only samples the total dust emission anyway) or there is full wavelength coverage in the infrared. If a young system is discovered in a gap accreting through a streamer and observations include the FIR, then the opacity could be measured directly using these models.

We have also assumed that the envelope is either present (when the system is embedded in the circumstellar disk) or absent (when the system is revealed in a deep gap), and that the system is azimuthally symmetric. Since the orbital timescale of the circumplanetary disk is much less than the infall timescale, any large asymmetries would be corrected. It is also possible for the envelope to maintain an azimuthal asymmetry from either spiral density waves and/or sub-Keplerian rotation in the background disk \citep{Cimerman2017, Krapp2022, Kuwahara2024}, but these effects are likely subdominant. 

%DIF >  In addition to assuming that the internal luminosity is small, this model also assumes that the luminosity in H$\alpha$ and the UV continuum is small compared to the photosphere luminosity. In combination, these assumptions imply that the photospheric luminosity is set by the accretion luminosity. While the H$\alpha$ luminosity is small compared to the accretion luminosity \citep{Thanathibodee2019, Aoyama2021}, if the accretion luminosity is instead mostly released in the UV continuum then the photospheric luminosity calculated here is an underestimate of the total luminosity. Additional mass accretion or a large internal luminosity would be necessary to account for this unaccounted-for luminosity. However, observations suggest that the UV continuum luminosity is small compared to the photospheric luminosity \citep{Finley2026}, so that the fundamental assumption relevant assumption in this work is that the internal luminosity is small compared to the accretion luminosity. 

{Finally, these models have neglected dust settling within the circumplanetary disk. If large dust is efficiently settled and most of the dust mass resides in large dust, then the SED is relatively unchanged, but the \qty{10}{\micro\meter}silicate feature becomes more prominent (see, e.g., \citealt{DAlessio2006}). The large dust dominates the opacity, so the continuum $\tau=1$ surface tends to follow the vertical distribution of the large dust. On the other hand, since the small dust has a significantly stronger silicate feature, the location of the \qty{10}{\micro\meter} $\tau=1$ surface is set by the vertical distribution of the small dust. As the large dust settles relative to the small dust, the location of these two surfaces diverges, and the temperature at the $\tau=1$ surface varies significantly at the \qty{10}{\micro\meter} feature. Settling of large dust therefore significantly strengthens the silicate feature. Detailed discussion of the effects of dust settling are left for future work. For a well-mixed disk, the silicate feature is generally weak, since the $\tau=1$ surfaces fall in a region where the temperature is relatively flat in $z$, so that the blackbody continuum and feature emission at the same temperature. The effects of dust settling will be especially significant in the FIR, so constraints using observations in the FIR will be especially sensitive to these effects. The effects of dust settling and grain growth will be discussed in future work. 
}

{This model also relies heavily on structural assumptions for the disk structure, the relative distributions of the dust and gas, and the dust size distribution (see Appendix \ref{sec:dustsize} for a discussion of the effect of maximum dust size). Since the observed flux is strongly dependent on both the system luminosity and the disk structure, changes to the assumptions such as invoking significant dust settling may change the mapping between observables and physical properties, which would then force the MCMC model to select different optimal parameters. These constraints are not absolute, but are subject to the assumptions of this particular model.}

Future work is necessary to investigate precisely when young protoplanets can be detected and to design survey strategies to find these objects (e.g.,  {\citealt{Szulagyi2018, Szulagyi2019, Krieger2022, Chen2022, Sun2024}} ). It will also be useful to consider the time-evolution of these planets and their accompanying radiative signatures as part of this process. Such a program would aid in the detection of more of these objects and improve our understanding of forming planets. 

With the advent of \qty{40}{\meter}-class telescopes, many more circumplanetary disks are likely to be detected in the near future. This work has provided an efficient and effective model for infrared SEDs that enable these observations to constrain the parameters of these protoplanets. Although observations with existing and planned instruments can provide useful constraints, the construction of an FIR instrument is necessary for continuum observations to jointly constrain the planet mass and mass accretion rate. The construction of such an instrument would provide a great benefit in exoplanet studies as more circumplanetary disks are detected.

\section*{Acknowledgments}

We thank Nuria Calvet, Maria Jose Colmenares, Gabriele Cugno, Michael Meyer, and William Meynardie for helpful conversations and advice. A.G.T. acknowledges support from the Fannie and John Hertz Foundation and the University of Michigan's Rackham Merit Fellowship Program. 
F.C.A. is supported in part by the Leinweber Institute for Theoretical Physics at the University of Michigan and by NSF grant No. 2508843. 

This work has made use of the Julia programming language \citep{Julia} and the plotting package Makie \citep{Makie}. 

\appendix

\section{Semianalytic Model Details}\label{app:SAMmath}

This section discusses the details of the calculation of the semianalytic disk model, which depends on the model parameters ($M_p$, $\dot{M}$, $a$, etc) and three structure parameters ($\dasp$, $q$, and $f_L$). 

These structure parameters can be easily calculated for a given \RADp model. The slope of the disk surface $\dasp$ is found by evaluating $z_{\rm surf}/R$ from $R_X$ to $R_C$ and taking the median value. The pole-emission luminosity ratio $\fL$ is found by integrating the SED when viewed from the system pole and dividing by the total luminosity of the system. The temperature slope $q$ is calculated as the best-fit power-law index to the surface temperature $T(z_{\rm surf})$ as a function of $R$. 

For an arbitrary set of model parameters, it is not practical to calculate a \RADp model at that point in order to find the structure parameters. Instead, the values of $\dasp$, $q$, and $\fL$ are found by interpolation. \RADp models have been calculated at a series of points in parameter space. The interpolation is accomplished using radial basis functions, with a multiquadratic kernel and an additional linear term. Outside of the domain encompassed by these ``training'' models, extrapolations are generally linear. 

\begin{figure}
    \centering
    \includegraphics{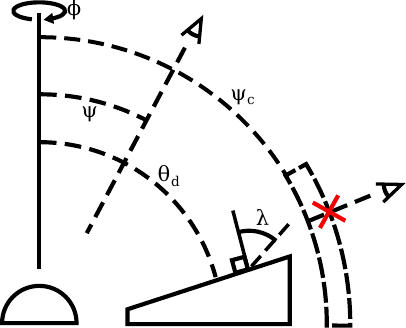}
    \caption{\textbf{Angles of the Disk System.} The various angles defined for the disk system. The planet (bottom left) and the disk are not to scale. The half-opening angle $\theta_d$, the viewing angle $\psi$, the critical viewing angle $\psi_c$, the azimuthal angle $\phi$, and the angle normal to the surface $\lambda$ are all shown.}
    \vspace{-10pt}
    \label{fig:angledig}
\end{figure}

\subsection{Disk Calculation}

Once the structure parameters have been calculated, the disk's temperature and radiative signatures are fully defined. The temperature of the disk $T_d$ is assumed to take the form 
\begin{equation}\label{eq:Tdpl}
    T_d=T_X\left(\frac{R}{R_X}\right)^{-q}\,.
\end{equation}
In general, the temperature distribution at the $\tau_d=1$ emission surface of the \RADp disks closely follow this power-law distribution. 

If the outer disk is optically thin, its temperature profile may not follow the same $q$ as the inner, optically thick disk. In an optically thin outer disk, the power-law index is set by a balance of local heating and radiative cooling, since all of the emitted radiation escapes. For a small disk cell, the emission luminosity is proportional to $L_{-}\propto\rho\kappa_PT^4$. If the disk does not absorb any radiation, the heating is exclusively due to viscous heating so that $L_+\propto\Gamma(R)\propto\rho T\Omega$. If $\kappa_\nu\propto\nu^{\gamma}$, then $\kappa_P\propto T^{\gamma}$ (see \citealt{Adams1985, Taylor2025}). Since $\Omega\propto R^{-3/2}$, setting $L_+\sim L_-$ shows that $T\propto R^{-3/(6+2\gamma)}$. Here, we assume $\gamma=1/2$ so that $T\propto R^{-3/7}$ in the outer, optically thin disk. The power law normalization is set so that the thick and thin disk have the same temperature at a radius $R_{\rm thin}$. 

The turnover point $R_{\rm thin}$ is assumed to be where the total optical depth of the disk is equal to one. For a given temperature normalization $T_X$ and radius $R_{\rm thin}$, the surface density can be calculated and $R_{\rm thin}$ checked. The value of $T_X$ depends on the value of $R_{\rm thin}$. For a given $T_X$, $R_{\rm thin}$ is then calculated via a simple binary search algorithm between $R_X$ and $R_C$ to find where $R_{\rm thin}$ can be placed such that the disk has $\tau=1$ at $R=R_{\rm thin}$. 

The only remaining undefined disk parameter is the temperature normalization $T_X$. This normalization is specified by $\fL$, which relates the emergent flux along the pole to the total luminosity of the system. The disk emits from three surfaces --- the inner wall, the outer wall, and the conical disk photosphere. When viewed along the pole, only the disk surface is visible, and so the disk must satisfy the constraint 
\begin{equation}\label{eq:Econ}
    \fL(L_d+L_p)=L_{d, \text{surf}}(\psi=0)+L_{p}(\psi=0)\,.
\end{equation}
% Here, $X_{\rm abs}$ is the fraction of the planet's luminosity that is absorbed and reprocessed by the disk. $L_{d, \text{surf}}$, $L_{d, \text{iw}}$, and $L_{d,\text{ow}}$ are the luminosity emitted by the disk surface, the inner wall of the disk, and the outer wall of the disk, respectively. 

First, the integrated SED of the planet along the pole is simply given by 
\begin{equation}
    L_p(\psi=0)=4\pi R_p^2\sigma T_p^4\,.
\end{equation}
Then, for a given point on the disk at $R$ with temperature $T$, the disk is modeled as a plane-parallel slab with an optical depth $\tau_\nu$ in the normal direction. If the viewing direction has an angle $\lambda$ relative to the disk normal, this point emits an intensity 
\begin{equation}
    I_\nu=B_\nu(T)[1-\exp(-\tau_{\nu,\perp}/\cos\lambda)]\,.
\end{equation}
For a surface area element $\de A_\perp$, the luminosity in this direction is given by 
\begin{equation}
    \de L_\nu=F_\nu\de A=\oint\cos\theta\,I_\nu\de\Omega\simeq\cos\lambda\, I_\nu\de A_\perp\,,
\end{equation}
where $\cos\lambda$ is introduced due to the projection of the disk. For a given viewing angle $\psi$ relative to the vertical and an azimuthal position $\phi$ relative to the viewing direction, the law of cosines provides that 
\begin{equation}
    \cos\lambda=\cos\psi\sin\theta_d - \cos\phi\sin\psi\cos\theta_d\,.
\end{equation}
However, the surface area element $\de A_\perp=\de A/\sin\theta_d$, where $\de A=R\,\de R\,\de\phi$ is the cylindrical area element. This additional factor is introduced by the slope of the disk photosphere relative to the cylindrical definite plane. The normal optical depth $\tau_{\nu,\perp}$ is related to the optical depth in the vertical direction $\tau_{\nu, v}$ by $\tau_{\nu,\perp}=\sin\theta_d\,\tau_{\nu, v}$. The vertical optical depth of the disk $\tau_{\nu,v}=\Sigma(R)\kappa_\nu/2$.

% It is important to note that this area element cannot be viewed from all viewing angles $\psi$. First, if $\lambda>\pi/2$, then the observer is viewing the system from below the disk surface. In addition, at significant viewing angles, the near side of the disk will block the viewing of the far side of the disk. This critical viewing angle $\psi_c$ depends on $R$, the azimuthal position angle $\phi$, and $f_d$. This critical angle is the angle with respect to the $z$-axis in the $x$-$z$ plane between the point on the disk defined by $R, \phi$ and the disk rim in the positive $x$ direction. If the viewing angle $\psi$ is greater than this angle, the edge of the disk will obscure this portion of the disk surface. The relevant disk edge must have the same $x$-coordinate as the disk surface point, height $f_dR_C$, and $x$-$y$ plane radius $R_C$. The critical angle can be found to be 
% \begin{equation}
%     \tan\psi_c=\frac{\sqrt{R_C^2-R^2\sin^2\phi}-R\cos\phi}{f_d(R_C-R)}\,.
% \end{equation}
% This critical angle defines the self-shadowing of the disk. Since the disk is a cone, requiring $\psi<\psi_c$ also forces $\lambda\leq\pi/2$. 

The luminosity from the disk surface at the system pole can then be found by integrating over the entire disk and all frequencies, so that 
\begin{equation}
\begin{split}\label{eq:Ldsurf}
    L_{d,\text{surf}}=4\pi\!\!\int\displaylimits_{R_X}^{R_C}\!\!\de R&\!\!\int\displaylimits_0^{2\pi}\!\!\de\phi\!\!\int_0^\infty\!\!\de\nu\sin\psi\,R\frac{\cos\lambda}{\sin\theta_d}B_\nu(T(R))\\
    &\times\left[1-\exp\left(-\frac{\tau_{\nu, v}\sin\theta_d}{\cos\lambda}\right)\right]\,.
\end{split}
\end{equation}
% If the disk is optically thick over all $R$, then the exponential term is very small and can be dismissed. In this case, the integral of $\cos\lambda\sin\psi\de\psi$ can be evaluated analytically to become 
% \begin{equation}
%     \int_0^{\psi_c}\!\!\de\psi\sin\psi\,\cos\lambda=\sin^2\psi_c-\psi_c\cot\theta_d\,\cos\phi+\sin\psi_c\cos\psi_c\cot\theta_d\cos\phi\,.
% \end{equation}
Note that if $\psi=0$ then $\cos\lambda=\sin\theta_d$, which significantly simplifies the integral above.
The wavelength integral can then be simplified to $\int_0^\infty B_\nu(T)\de\nu=\sigma T^4(R)/\pi$ and 
\begin{equation}
    \int\displaylimits_0^\infty\!\! B_\nu(T)\exp(-\kappa_\nu N)\,\de\nu=\frac{1}{\pi}\sigma T^4\alpha_P(T,N)\,,
\end{equation}
by definition of the effective absorption coefficient $\alpha_P$. 
For a given $T_X$, $L_{d,\text{surf}}$ is fully specified. A bisection root-finding algorithm is then used to find the $T_X$ that satisfies Eq. \eqref{eq:Econ}. 

The inner wall of the disk is not visible when viewed directly from above, but has a significant impact on the disk SEDs. The inner wall is necessarily optically thick, and each point on the inner wall is assumed to have the same temperature
\begin{equation}
    T_{\rm wall}^4=T_X^4+T_p^4\left(\frac{R_p}{R_X}\right)^2\,.
\end{equation}
Here, $T_p$ is the temperature of the planet and the rightmost term accounts for the heating of the inner wall by the planet. Since the inner wall has a constant temperature, the total luminosity emitted in a given direction is 
\begin{equation}
    L_{d,\text{iw}}=-8f_d R_X^2\sigma T_{\rm wall}^4\!\!\int\displaylimits_0^{\pi/2}\!\!\de\psi\!\!\int\displaylimits_{\pi/2}^{3\pi/2}\!\!\de\phi\sin^2\psi\,\cos\phi\,.
\end{equation}
The integral of $\cos\phi$ from $\pi/2$ to $3\pi/2$ is $-2$, so that
\begin{equation}\label{eq:Ldiw}
    L_{d,\text{iw}}=16\dasp R_X^2\sigma T_{\rm wall}^4\!\!\int\displaylimits_0^{\pi/2}\!\!\de\psi\sin^2\psi\,.
\end{equation}

While this methodology generally accurately finds the temperature $T_X$, it is not guaranteed to conserve energy. Without some additional effect, integrating $L_p(\psi)+L_d(\psi)$ over all solid angle is significantly larger than $L_p+L_d$. In fact, the planet, inner wall, and disk surface are shadowed by the disk itself. At steep viewing angles, only the outer wall is visible. The luminosity of the outer wall is always very low and so is generally ignored. At some point in between, there is a transition. For convenience, the transition will be modeled as a step function in $\psi$ with a cutoff of $\psi_c$, so that the entire system is visible for $\psi<\psi_c$ and only the outer wall is visible for $\psi>\psi_c$. The exact value of $\psi_c$ is chosen so that the energy of the system is conserved. 

\subsection{Envelope Calculations}

The radiation from the disk and planet are then further attenuated by the circumplanetary envelope. The total energy absorbed and re-emitted by the envelope is the difference between the total luminosity and the planet and disk radiation that escapes the envelope. 

For a given frequency-dependent opacity $\kappa_\nu$ and a line-of-sight column density $N_{\rm col}$, the escaping radiation is reduced by a factor of $\exp(-\kappa_\nu N_{\rm col})$. The radiation absorbed at a given viewing angle can then be integrated over all viewing angles to find the total energy deposited in the envelope. When viewing the planet at an angle $\psi$, the column density is given by 
\begin{equation}
    N_{p,\rm col}=\int\displaylimits_{0}^{R_H}\!\!\rho(r,\psi)\de r
\end{equation}
and the total escaping radiation is 
\begin{equation}\label{eq:Lpescsimp}
    L_{p,\rm esc}=\int\displaylimits_0^{\psi_c}\!\!\de\psi\!\!\int\displaylimits_0^\infty\!\!\de\nu\,4\pi^2R_p^2\sin\psi \,B_\nu(T_p)\exp(-\kappa_\nu N_{p, \rm col})\,.
\end{equation}
Recall that the planet cannot be seen for $\psi>\psi_c$. The integral $\de\nu$ can be simplified by calculating the effective attenuation constant $\alpha_P$, which is defined such that 
\begin{equation}
\begin{split}
    \alpha_P[T,N_{\rm col}]&\equiv\int\displaylimits_0^\infty\!\!B_\nu(T)\exp(-\kappa_\nu N_{\rm col})\,\de\nu\,\bigg/\!\!\int\displaylimits_0^\infty\!\!B_\nu(T)\,\de\nu\\
    &=\frac{\pi}{\sigma T^4}\int\displaylimits_0^\infty\!\!B_\nu(T)\exp(-\kappa_\nu N_{\rm col})\,\de\nu
\end{split}
\end{equation}
Eq. \eqref{eq:Lpescsimp} then becomes 
\begin{equation}
    L_{p,\rm esc}=4\pi R_p^2\sigma T_p^4\int\displaylimits_0^{\psi_c}\!\!\de\psi\sin\psi\,\alpha_P[T_p,N_{p,\rm col}(\psi)]\,.
\end{equation}

In addition, the inner wall of the disk is attenuated by the same number density as the planet, so that the escaping radiation from the inner wall takes the form 
\begin{equation}
    L_{d, \rm iw, esc}=16\dasp R_X^2\sigma T_{\rm wall}^4\int\displaylimits_0^{\psi_c}\!\!\sin^2\psi\,\alpha_P[T_{\rm wall}, N_{p,\rm col}(\psi)]\,.
\end{equation}

The surface of the disk is attenuated by a remarkably different column density. This column density is found by integrating along a ray from the point on the disk surface to the Hill sphere. The relevant ray is always in the same $x$-$z$ plane as the disk point and has a polar angle $\psi$. The distance along the ray can be characterized by a parameter $s$ so that the ray crosses the disk surface at $s=0$ and the Hill sphere at $s=1$. If the disk point is at a cylindrical radius $R$ and an angle $\phi$ relative to the $x$-axis, the distance between the endpoints $d$ and the position along the ray $r$, $\mu$ are given by 
\begin{subequations}
\begin{align}
    \begin{split} d = \Big(&R_H^2-R^2(1-\sin^2\psi\cos^2\phi)\\
    &-r\sin\psi\cos\phi\Big)^{1/2}\,; \end{split}\\
    \begin{split} r &=\bigg(s^2R_H^2+(1-s)^2 R^2(1+\dasp^2)\\
        &-2s(1-s)R\Big((R_H^2-d^2\cos^2\psi)\cos\phi\\
        &+dR_H\cos\psi\,\dasp\Big)^{1/2}\bigg)^{1/2}\,;\end{split}\\
    \mu &= \frac{sd\cos\psi+(1-s)\dasp R}{r}\,.
\end{align}
\end{subequations}
The column density between the observer and an arbitrary location on the disk surface $N_{d, \rm col}$ can then be found by integrating $d\,\rho(r[s],\mu[s])\de s$ along the ray. The radiation that escapes from the disk surface is then found by integrating Eq. \eqref{eq:Ldsurf} over the entire disk surface, with the additional attenuation factor of $\exp(-\kappa_\nu N_{d, \rm col}$ in the integrand. This function is then integrated again with a factor of $4\pi\sin\psi \,\de\psi$ to account for all solid angles. Recall that the disk surface cannot be seen for $\psi>\psi_c$, which sets the limits of this integral. 

The total energy that must be emitted by the envelope is then given by 
\begin{equation}\label{eq:Ledef}
    L_e=L_p+L_d-L_{p,\rm esc}-L_{d,\rm esc}\,.
\end{equation}
For a given parcel of envelope with a density $\rho$ and temperature $T$, the total energy that escapes in a direction $\psi$ is given by 
\begin{equation}
    L_{e,\nu}(\psi)=\rho\kappa_\nu r^2\sin\theta \,B_\nu(T)\,\de\theta\,\de r\,\de\phi\,.
\end{equation}
The total luminosity can be found by integrating $L_{e,\nu}(\psi)$ by solid angle $\de\Omega=4\pi\sin\psi\,\de\psi$ and frequency $\de\nu$. Note that integrating $\kappa_\nu B_\nu(T)\de\nu=\kappa_P\sigma T^4/\pi=b_\kappa \sigma T^{4+\gamma}/\pi$, by the definition of the Planck mean opacity $\kappa_P$. Now, since the envelope is optically thin to its own radiation, the temperature of the envelope scales as \citep{Adams1985}
\begin{equation}
    T_e(r)=T_C\left(\frac{r}{R_C}\right)^{-2/(4+\gamma)}\,.
\end{equation}

As discussed briefly in the main body of the text, the envelope is divided into three regions, each of which emit in different ways. The first region is situated in the interior of the disk. In this region, the temperature takes the power-law form in radius and the region is only visible for $\psi<\psi_c$. For $\psi<\psi_c$, any point in this cone is visible. This region extends from an inner edge at $r=R_X\csc\psi_c$ to an outer edge at $r=R_C\csc\psi_c$, where the $\csc\psi_c$ terms account for the fact that the disk walls have a fixed cylindrical radius, not a fixed spherical radius. The total luminosity of this first region is then given by 
\begin{equation}
    L_{e1}=8\pi(1-\cos\psi_c)\int\displaylimits_{R_X\csc\psi_c}^{R_C\csc\psi_c}\de r\!\!\int\displaylimits_{0}^{\psi_c}\!\!\de\theta\, \sin\theta\,T_C^{4+\gamma}R_C^2\rho(r,\theta)\,.
\end{equation}
The first term in this equation comes from the integral $\de\psi$ and the $T_e^{4+\gamma}$ neatly cancels out the $r^2$ term from the $\de r$. The result is an integrand with $T_C$ as a constant, which can then be moved through the integral. 

The second region extends from $R_C\csc\psi_c$ to the Hill sphere at $r=R_H$. In angular space, it too is limited to $\theta<\psi_c$, but this region is visible from any viewing angle. The temperature of the second region has the same power-law index as the first. Therefore, the luminosity of the second region is given by
\begin{equation}
    L_{e2}=8\pi\int\displaylimits_{R_C\csc\psi_c}^{R_H}\de r\!\!\int\displaylimits_{0}^{\psi_c}\!\!\de\theta\, \sin\theta\,T_C^{4+\gamma}R_C^2\rho(r,\theta)\,.
\end{equation}
Since the $\de\psi$ integral has an upper limit of $\pi/2$, the $1-\cos\psi_c$ term does not appear here. 

Finally, the third region is limited to the remaining region of the envelope space --- $r=R_C\csc\psi_c$ to $r=R_H$ and $\theta>\psi_c$. Like the second region, it is visible from any viewing position. However, the temperature of the third region is different from that of the other two. Since this region is shadowed by the circumplanetary disk, it is not heated by the central planet. Instead, the temperature of this region is set to $T_d(R_C)$, the temperature of the outer edge of the disk. The luminosity of this region is given by 
\begin{equation}
    L_{e3}=8\pi\sigma b_\kappa T_d(R_C)^{4+\gamma}\!\!\!\!\!\!\!\!\int\displaylimits_{R_C\csc\psi_c}^{R_H}\!\!\!\!\!\!\de r\!\!\int\displaylimits_{\psi_c}^{\pi/2}\!\!\de\theta\, \sin\theta\,r^2\rho(r,\theta)\,.
\end{equation}

The total luminosity of the envelope must be the sum of these three components. Helpfully, $L_{e1}$ and $L_{e2}$ can be written as an integral times $T_C^{4+\gamma}$. For $L_e$ given by Eq. \eqref{eq:Ledef}, a simple algebraic rearrangement can solve for $T_C$ and fully define the temperature. 

These calculations collectively precisely specify the temperature, structure, and radiative signatures of the circumplanetary disk system {. 
}

\begin{figure}
    \centering
    \includegraphics{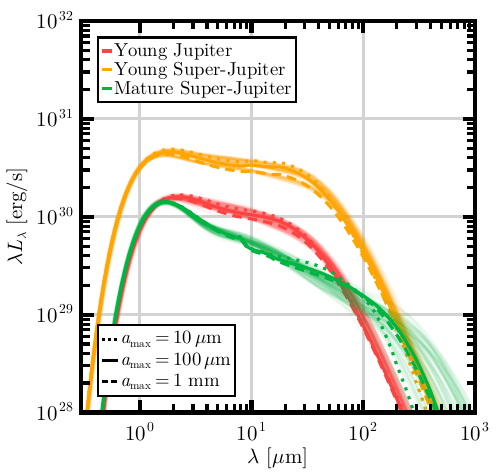}
    \vspace{-10pt}
    \caption{{\textbf{SED Effects of Maximum Dust Size.} The SEDs as calculated by \RADp for the three fiducial models assuming a maximum dust size of \qty{10}{\micro\meter} (dotted), \qty{100}{\micro\meter} (solid), and \qty{1}{\milli\meter} (dashed). The envelope is not included in any of these models. The SED posterior from fitting the SAM to these models are also shown. The dust size has little impact on the SEDs, except for the mature super-Jupiter at long wavelengths where the disk can become optically thin. The parameters returned by MCMC fits to the SEDs differ by factors of a few and are within the uncertainty envelope.}}
    \vspace{-10pt}
    \label{fig:dustsizefig}
\end{figure}

\section{{Dust Size Effects}}\label{sec:dustsize}

{This Appendix discusses the effect of the maximum dust size on the fiducial models. Fig. \ref{fig:dustsizefig} shows the SEDs for the three fiducial models introduced in Sec. \ref{sec:synthobv}, where the dust is set to have a maximum size of \qty{10}{\micro\meter}, \qty{100}{\micro\meter} (the default value) or \qty{1}{\milli\meter}. Since these disks are optically thick throughout, the change in the maximum dust size has little impact on the SEDs. The exception is the long wavelengths in the mature super-Jupiter model, where the disk is borderline optically thin and the maximum dust size has a minor effect. Although the SAM assumes a maximum dust size of \qty{100}{\micro\meter}, the similarity in the SEDs means that the model is still able to accurately fit the synthetic observations. The returned model parameters are within $\sim$1$\sigma$ of the correct value, with no clear bias due to changing dust sizes. The effect of the maximum dust size is thus minor, especially in the NIR and MIR where the circumplanetary disk is optically thick and the SEDs are identical} . 

\bibliography{main}{}
\bibliographystyle{aasjournal}

\end{document}